\documentclass[aps,nofootinbib,floatfix,showpacs,preprintnumbers,twocolumn]{revtex4} %eqsecnum,
\usepackage{graphicx}
\usepackage{bm}
\usepackage{mathrsfs}
\usepackage{float}
\usepackage{tabularx} 
\usepackage{amsmath}
\usepackage{epstopdf}
\usepackage{multirow}
\usepackage{color}
\usepackage{booktabs}
\usepackage{hyperref}
\usepackage{tabularx} % in the preamble
\input epsf

\def\lsim{\mathrel{\raise.3ex\hbox{$<$\kern-.75em\lower1ex\hbox{$\sim$}}}}
\def\gsim{\mathrel{\raise.3ex\hbox{$>$\kern-.75em\lower1ex\hbox{$\sim$}}}}

\def\cmm2{{\,\rm cm^{-2}}}
\def\cm2{{\,{\rm cm}^2}}
\def\cmm3{{\,{\rm cm}^{-3}}}
\def\gcmm3{{\,{\rm g\,cm^{-3}}}}

\def\fun#1#2{\lower3.6pt\vbox{\baselineskip0pt\lineskip.9pt
  \ialign{$\mathsurround=0pt#1\hfil##\hfil$\crcr#2\crcr\sim\crcr}}}

\def\etal{{\it et al.}}

\def\be{\begin{equation}}
\def\ee{\end{equation}}
\def\bea{\begin{eqnarray}}
\def\eea{\end{eqnarray}}

\newcommand{\vesc}{v_{\rm esc}}

\newcommand{\vmin}{v_\textrm{min}}

\newcommand{\ra}[1]{\renewcommand{\arraystretch}{#1}}

\begin{document}

\title{Dark matter astrophysical uncertainties and the neutrino floor}

\author{Ciaran A. J. O'Hare }\email{ciaran.ohare@nottingham.ac.uk} \affiliation{School of Physics and Astronomy, University of Nottingham, University Park, Nottingham, NG7 2RD, UK}

\date{\today}
\smallskip
\begin{abstract}
The search for weakly interacting massive particles (WIMPs) by direct detection faces an encroaching background due to coherent neutrino-nucleus scattering. For a given WIMP mass the cross section at which neutrinos constitute a dominant background is dependent on the uncertainty on the flux of each neutrino source from either the Sun, supernovae or atmospheric cosmic ray collisions. However there are also considerable uncertainties with regard to the astrophysical ingredients to the predicted WIMP signal. Uncertainties in the velocity of the Sun with respect to the Milky Way dark matter halo, the local density of WIMPs, and the shape of the local WIMP speed distribution all have an effect on the expected event rate in direct detection experiments and hence will change the region of the WIMP parameter space for which neutrinos are a significant background. In this work we extend the WIMP+neutrino analysis to account for the uncertainty in the astrophysics dependence of the WIMP signal. We show the effect of uncertainties on projected discovery limits with an emphasis on low WIMP masses (less than 10 GeV) when Solar neutrino backgrounds are most important. We find that accounting for astrophysical uncertainties changes the shape of the neutrino floor as a function of WIMP mass but also causes it to appear at cross sections up to an order of magnitude larger, extremely close to existing experimental limits - indicating that neutrino backgrounds will become an issue sooner than previously thought. We also explore how neutrinos hinder the estimation of WIMP parameters and how astrophysical uncertainties impact the discrimination of WIMPs and neutrinos with the use of their respective time dependencies.
\end{abstract}
\pacs{95.35.+d; 95.85.Pw}
\maketitle

\section{Introduction}
The nature and detection of dark matter is one of the most pressing unsolved problems in modern physics. A myriad of cosmological observations indicate that $\sim$30\% of the energy density of the Universe must be comprised of a cold and non-baryonic component, yet the particle content of this dark matter remains unknown~\cite{Ade:2015xua}. The most promising method of detecting dark matter in the laboratory is the search for their keV-scale nuclear recoils produced in elastic scattering events between dark matter particles in the Milky Way halo and target nuclei~\cite{Goodman:1984dc}. This method of detection is possible if dark matter is in the form of a Weakly Interacting Massive Particle (WIMP). These particles are a popular and well motivated candidate which appear in extensions to the Standard Model such as Supersymmetry, and freeze out in the early Universe with an abundance that matches cosmological observations (for reviews see e.g., Refs.~\cite{Jungman:1995df,Bertone:2004pz}). 

Neutrinos are also weakly interacting particles. It is known that they must, too, elastically scatter off the same target nuclei of dark matter detectors. As a result of neutrinos being impossible to shield against, they are regarded as the ultimate background to experimental searches for WIMPs~\cite{Monroe:2007xp}. Current experiments such as Xenon100~\cite{Aprile:2012nq}, LUX~\cite{Akerib:2013tjd}, and CDMS~\cite{Agnese:2014aze}, which can probe to spin-independent (SI) WIMP-nucleon cross sections of the order $\sigma_{\chi-n} \approx  10^{-44}-10^{-45} \, {\rm cm}^2$, are not yet sensitive to the expected neutrino background (for a recent review of direct detection experiments see e.g., Ref.~\cite{Undagoitia:2015gya}). However as the sensitivity of experiments increases with the next generation of ton-scale (and beyond) detectors such as Xenon1T~\cite{Aprile:2015uzo} and LZ~\cite{Akerib:2015cja}, the neutrino background will begin to become important~\cite{Cabrera:1984rr,Monroe:2007xp,Strigari:2009bq,Gutlein:2010tq}. The limiting WIMP parameters at which this occurs is known as the neutrino floor and is caused by a close similarity between the recoil energies and rates of WIMPs of certain masses and cross sections~\cite{Monroe:2007xp,Strigari:2009bq,Gutlein:2010tq,neutrinoBillard}. For example in a Xenon detector the recoil energy spectrum of a WIMP with mass $m_{\chi}= 6 \, {\rm GeV}$ and SI cross section $\sigma_{\chi-n} \sim 5 \times 10^{-45} \, {\rm cm}^2$ very closely matches that of ${}^{8}\rm{B}$ Solar neutrinos~\cite{Strigari:2009bq}.  

The neutrino floor is not however the true final limit to direct detection. As initially shown by Ruppin~\etal~\cite{neutrinoRuppin}, the differences in the tails of the recoil energy distributions of WIMPs and neutrinos allow the ``floor'' to be overcome albeit with very large numbers of events ($>\mathcal{O}(1000)$). Furthermore other studies have shown that the neutrino background can be mitigated with the use of annual modulation effects~\cite{Davis:2014ama}, direction dependence~\cite{Grothaus:2014hja, O'Hare:2015mda} or complementarity between multiple target nuclei~\cite{neutrinoRuppin}. A recent work by Dent \etal~\cite{Dent:2016iht} made use of the non-relativistic effective field theory formalism which posits additional operators to describe the nuclear response to a WIMP interaction. They found that for many of these additional operators, which induce significantly different nuclear recoil energy spectra, the neutrino bound is much weaker or not present. However the limits calculated by these studies all depend on the choice of astrophysical input, hence if an accurate prediction is to be made about when future experiments will be affected by the neutrino background at a statistically significant level, we must first establish the extent to which the uncertainty in the astrophysical input plays a role. 

The phase space structure of the Milky Way halo in the region of the Solar System is an uncertain and much debated topic~\cite{Persic:1995ru,Klypin:2001xu,Strigari:2013iaa}. Most direct detection analyses use an isotropic and isothermal spherically symmetric assumption for the Galactic halo known as the standard halo model (SHM). This model gives rise to an isotropic Maxwellian velocity distribution for which there is an analytic form for the WIMP event rate. However substantial evidence from N-body and hydrodynamic simulations indicate that the velocity structure is likely to contain significant deviations from this simple Maxwellian form~\cite{Vogelsberger:2008qb, Kuhlen:2009vh, Mao:2012hf}. In extreme cases there is some evidence that the velocity distribution may contain highly anisotropic features such as tidal streams~\cite{Freese:2003tt,Purcell:2012sh}, a dark disk~\cite{Ling:2009cn,Bruch:2008rx,Read:2008fh,Read:2009iv} and debris flows~\cite{Lisanti:2011as,Kuhlen:2012fz} which have been shown to have a noticeable effect on direct detection signals~\cite{Savage:2006qr,Lee:2012pf,O'Hare:2014oxa}. To add further complication, the astrophysical parameters not related to the velocity distribution directly such as the laboratory velocity $\textbf{v}_\textrm{lab}$ and local density $\rho_0$ are also subject to a high degree of uncertainty~\cite{McCabe:2010zh}. Particularly in the case of the local density this is a source of concern as it is degenerate with the WIMP-nucleon cross section. Attempts have been made however to account for astrophysical uncertainties in direct detection analysis by calculating halo-independent discovery limits~\cite{Frandsen:2011gi,Feldstein:2014gza,Anderson:2015xaa}. In this work we will incorporate neutrino backgrounds into an analysis which will take into account these astrophysical uncertainties.

The outline of this paper is as follows, in Sec.~\ref{sec:theory} we present the nuclear recoil event rates in the case of WIMP and neutrino elastic scattering. Then in Sec.~\ref{sec:nufloor} we describe the mock detector and outline the profile likelihood ratio test employed in order to calculate the neutrino floor. In Sec.~\ref{sec:astro} we describe the various sources of astrophysical uncertainty and how estimates on each individual parameter and different speed distributions change the neutrino floor. Section~\ref{sec:parcon} deals with the effect of the same uncertainties on the parameter constraints possible in the event that a WIMP signal is detected. In Sec.~\ref{sec:future} we discuss the potential for the upcoming generation of direct detection experiments to detect coherent neutrino-nucleus scattering, and in Sec.~\ref{sec:time} we show how it is possible to go beyond the putative neutrino floor by including the time dependence of the WIMP and Solar neutrino event rates; in both Sections we emphasise the importance of understanding astrophysical uncertainties. Finally in Sec.~\ref{sec:conc} we summarise our results and conclude.

\begin{figure*}[t]
\begin{center}
\includegraphics[trim = 0mm 0mm 0mm 0mm, clip,width=0.49 \textwidth,angle=0]{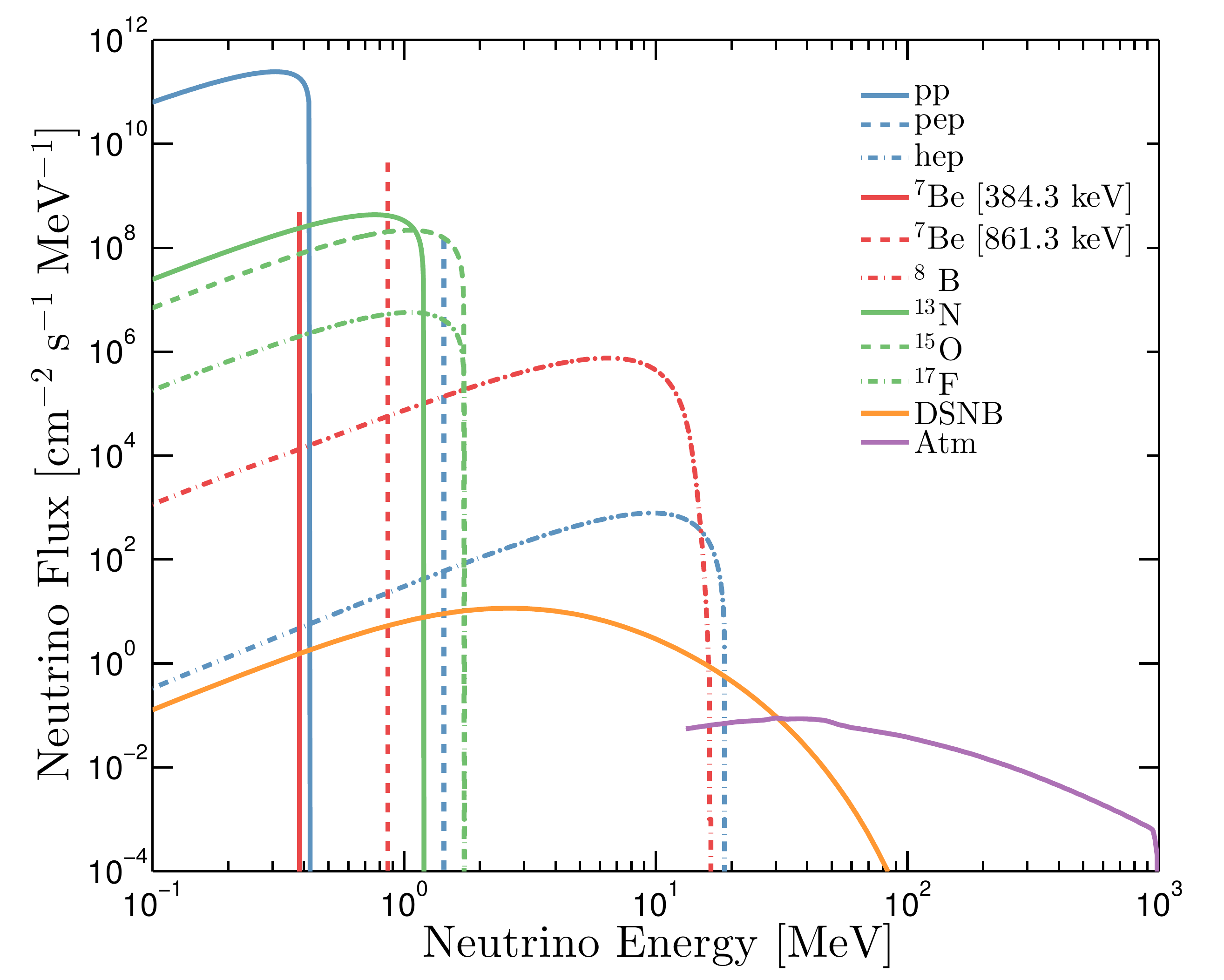}
\includegraphics[trim = 0mm 0mm 0mm 0mm, clip,width=0.49 \textwidth,angle=0]{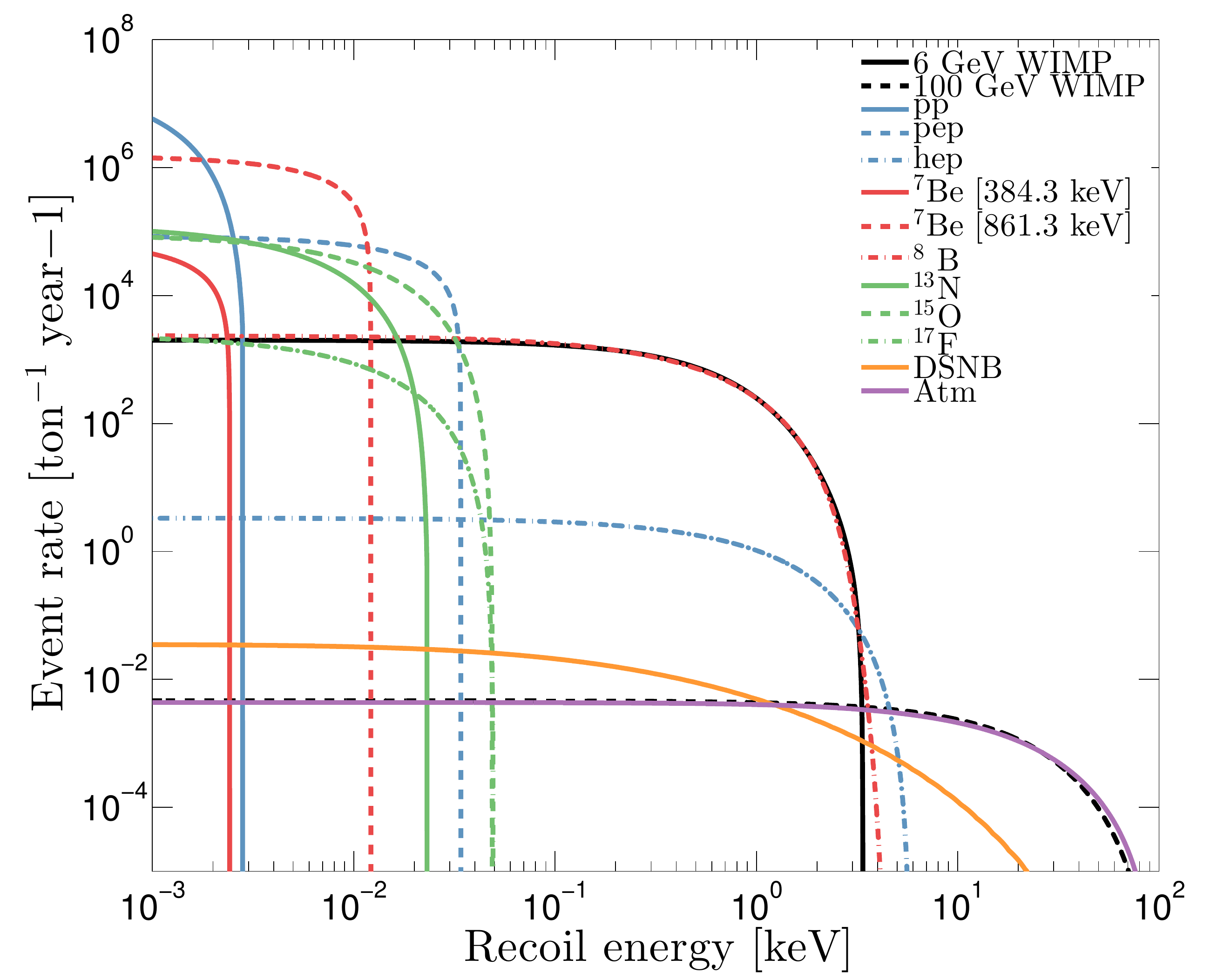}
\caption{{\bf Left:} Neutrino energy spectra that are backgrounds to direct detection experiments: Solar, atmospheric, and the diffuse supernova background. The Solar neutrino fluxes are normalised to the high metallicity standard Solar model. The atmospheric neutrino spectrum is the sum of contributions from electrons, anti-electrons, muons and anti-muons. The diffuse supernova background is the sum of three different neutrino temperatures, 3, 5 and 8 MeV. {\bf Right:} Xenon scattering event rate as a function of recoil energy for each type of neutrino as well as a 6 GeV WIMP with $\sigma_{\chi-n} = 5 \times 10^{-45}$~cm$^2$ (solid black line) and a 100 GeV WIMP with $\sigma_{\chi-n} = 2.5 \times 10^{-49}$~cm$^2$ (dashed black line) to show how they overlap with $^8$B and atmospheric neutrino induced recoils respectively.} 
\label{fig:neutrino_flux}
\end{center}
\end{figure*}

\section{Event rates}\label{sec:theory}
\subsection{WIMPs}
The elastic scattering event rate as a function of recoil energy and time assuming spin-independent interactions with identical couplings to protons and neutrons, is given by~\cite{Lewin:1995rx},
\begin{equation}
\frac{\mathrm{d}R_\chi}{\mathrm{d}E_r} = \frac{\rho_0\sigma_{\chi-n}}{2 m_{\chi}\mu^2_{\chi n}} A^2 F^2(E_r)g(\vmin,t),
\label{eq:WIMPrate}
\end{equation}
where $m_{\chi}$ is the WIMP mass, $\mu_{\chi n}$ is the WIMP-nucleon reduced mass, $\rho_0$ is the local dark matter density, $A$ is the mass number of the target and $\sigma_{\chi-n}$ is the WIMP-nucleon cross section. The function $F(E_r)$ is the form factor of the nucleus which only has a noticeable effect at large WIMP masses. In this work we will primarily be concerned with light WIMPs where understanding the neutrino background is most important, hence we will assume the standard Helm form factor for simplicity. Finally, $g(\vmin,t)$ is the mean inverse speed which is calculated by integrating the velocity distribution $f(\textbf{v})$ from $v_{\text{min}} = \sqrt{2 m_N E}/{2 \mu_{\chi n}}$ - the minimum WIMP speed required to produce a nuclear recoil of energy $E_r$, 
\begin{equation}
g(\vmin,t) = \int_{\vmin}^{\infty} \frac{f(\textbf{v} + \textbf{v}_\textrm{lab}(t))}{v} \, \textrm{d}^3 v \, .
\end{equation}  
The lab velocity, ${\bf v}_{\rm lab}$, is the velocity of the observer relative to the rest frame of the Galaxy and if taken to be time dependent is responsible for the annual modulation of the event rate~\cite{annual}. 

\subsection{Neutrinos}
For the keV-scale nuclear recoils relevent for WIMP detection the most important background to consider from neutrinos is due to coherent neutrino-nucleus scattering (CNS). We will ignore neutrino-electron elastic scattering, which for direct detection experiments has a very small event rate (from $pp$ neutrinos) and only slightly adjusts the neutrino floor for WIMP masses larger than $\sim$100 GeV~\cite{neutrinoBillard}. Despite the fact that CNS is yet to be observed, the rate is fully predicted by the Standard Model~\cite{Freedman:1973yd}. The differential cross section as a function of nuclear recoil energy ($E_r$) and neutrino energy ($E_\nu$) is,
\begin{equation}
  \frac{\textrm{d} \sigma}{\textrm{d}E_r}(E_r,E_\nu) = \frac{G_F^2}{4 \pi} Q_W m_N \left(1-\frac{m_N E_r}{2 E_\nu^2} \right) F^2(E_r) \,,
\end{equation}
where $Q_W = \mathcal{N} - (1-4\sin^2\theta_W) \mathcal{Z}$ is the weak nuclear hypercharge of a nucleus with $\mathcal{N}$ neutrons and $\mathcal{Z}$ protons, $G_F$ is the Fermi coupling constant, $\theta_W$ is the weak mixing angle and $m_N$ is the nucleus mass. The event rate per unit mass, as a function of the recoil energy is found by integrating the differential cross section with the neutrino flux from $E_\nu^{\rm min} = \sqrt{m_NE_r/2}$, which is the minimum neutrino energy required to generate a nuclear recoil with energy $E_r$,
\begin{equation}
  \frac{\textrm{d} R_\nu}{\textrm{d}E_r} = \int_{E_\nu^{\rm min}} \frac{\textrm{d} \sigma}{\textrm{d}E_r}\frac{\textrm{d}\Phi}{\textrm{d}E_\nu} \textrm{d}E_\nu \,.
\end{equation}
The neutrino flux $\Phi$ is the sum of multiple different components each with different individual energies and uncertainties. The relevent contributions to the neutrino background for WIMP searches are displayed in the left hand of Fig.~\ref{fig:neutrino_flux} with uncertainties listed in Table~\ref{tab:neutrino}. In fact with advances in technology currently underway~\cite{Mirabolfathi:2015pha} it will be possible for direct detection experiments to make competitive measurements of these neutrino fluxes~\cite{Strigari:2016ztv} and even constrain new physics such as the existence of sterile neutrinos~\cite{Billard:2014yka} or new interactions between neutrinos and nuclei or electrons~\cite{Cerdeno:2016sfi}.

Neutrinos from various fusion reactions in the interior of the Sun dominate the low energy-high flux regime and are the dominant background for direct detection with a total flux at Earth of around $6.5\times10^{11}$~cm$^{-2}$~s$^{-1}$~\cite{Robertson:2012ib,Antonelli:2012qu}. Neutrinos from the initial $pp$ reaction make up 86\% of all Solar neutrinos and have been detected most recently by the Borexino experiment, determining the flux with an uncertainty of $\sim 1$\% ~\cite{Bellini:2014uqa}. However for the remaining Solar neutrinos the theoretical uncertainties in the fluxes are as large as or larger than the measurement uncertainty and rely on an assumption of a Solar model for their calculation. In this work we assume the high metallicity Standard Solar Model (SSM)~\cite{Robertson:2012ib} which is the model most consistent with existing Solar data. Due to their relatively low energies, Solar neutrinos will influence the detection of WIMPs with masses less than 10 GeV. 

For WIMP masses between 10 and 30 GeV, the neutrino floor is caused by the sub-dominant diffuse supernova neutrino background (DSNB), the sum total of all neutrinos emitted from supernovae over the history of the Universe. The background flux is calculated by performing a line of sight integral of the spectrum of neutrinos from a single supernova with the rate density of core-collapse supernovae as a function of redshift. See Ref.~\cite{Beacom:2010kk} for the full calculation of the predicted DSNB. The total flux  of the DSNB is considerably smaller than for Solar neutrinos, around 86~cm$^{-2}$~s$^{-1}$, however it is an important background to consider as it extends to a higher energy range not occupied by Solar neutrinos. The calculated spectra have a Fermi-Dirac form with temperatures in the range 3 to 8 MeV. In this study we use a DSNB flux which is the sum of 3 temperatures: 3 and 5 MeV for electron and anti-electron neutrinos respectively, and 8 MeV for the sum of the remaining neutrino flavours. There are considerable theoretical uncertainties in this calculation, hence we will take a large systematic uncertainty of 50\% on the total flux of DSNB neutrinos~\cite{Beacom:2010kk}.

The final type of neutrino remaining to be discussed are those from the atmosphere which provide the main neutrino background for WIMP masses above 100 GeV. These neutrinos occupy the high energy and low flux regime and will limit the sensitivity of experiments to spin-independent cross sections below around $10^{-48}$cm$^{2}$~\cite{Strigari:2009bq,neutrinoBillard,neutrinoRuppin}. The flux of atmospheric neutrinos with energies less than 100 MeV is difficulat to measure as well as predict theoretically~\cite{Battistoni:2005pd,Honda:2011nf,Battistoni:2002ew} although the expected flux is around 11~cm$^{-2}$~s$^{-1}$. In this work we use a calculation that is a sum of the contributions from electron, anti-electron, muon and anti-muon neutrinos and place a $\sim 20\%$ uncertainty on the total flux~\cite{Honda:2011nf}.

In the right hand panel of Fig.~\ref{fig:neutrino_flux} we show the recoil energy spectrum for each neutrino type scattering with a Xenon target. In addition we show the recoil energy spectra for two example WIMPs with masses of 6 GeV and 100 GeV. This is to demonstrate the similarity that certain WIMP masses have with individual neutrino sources. This overlapping between WIMP and neutrino event rates is the reason why neutrinos limit WIMP discovery. For cross sections below the neutrino floor, an experiment which observes the excess in the number of observed events over the expected background cannot determine whether these events were due to a WIMP interaction or a slightly larger value of neutrino flux due to the systematic uncertainty. Hence the neutrino floor limit divides the WIMP parameter space into cross sections which induce enough events to be significant over the systematic uncertainty on the neutrino background and those which do not.

\begin{table}[t]
\begin{ruledtabular}
\begin{tabular}{cccc}
$\mathbf{\nu}$ \bf{type} & $\mathbf{E_{\nu}^{\rm{max}}}$ \bf{(MeV)} & $\mathbf{E_{r_{\rm{Xe}}}^{\rm{max}}}$ \bf{(keV)} & $\mathbf{\nu}$ \bf{flux}\\
 & & & $\mathbf{(\rm{cm^{-2} \, s^{-1}})}$\\
\hline
$pp$ & 0.42341 & $2.94\times 10^{-3}$ & $\left(5.98\pm 0.006 \right) \times 10^{10}$\\
${}^{7}\rm{Be}_{384.3}$ & 0.3843 & $2.42\times10^{-3}$ & $\left( 4.84\pm 0.48 \right) \times 10^8$\\
${}^{7}\rm{Be}_{861.3}$ & 0.8613 & 0.0122 & $\left( 4.35\pm 0.35 \right) \times 10^9$\\
$pep$ & 1.440 & 0.0340 & $\left( 1.44\pm 0.012\right) \times 10^8$\\
${}^{13}\rm{N}$ & 1.199 & 0.02356 & $\left(2.97\pm 0.14\right) \times 10^8 $\\
${}^{15}\rm{O}$ & 1.732 & 0.04917 & $\left(2.23\pm 0.15\right) \times 10^8$\\
${}^{17}\rm{F}$ & 1.740 & 0.04962 & $\left(5.52\pm 0.17 \right) \times 10^6$\\
${}^{8}\rm{B}$ & 16.360 & 4.494 & $\left(5.58\pm 0.14\right) \times 10^6$\\
$hep$ & 18.784 & 5.7817 & $\left( 8.04\pm 1.30 \right) \times 10^3$\\
DSNB & 91.201 & 136.1 & $ 85.5\pm 42.7$\\
Atm. & 981.748 & $15.55\times 10^3$ & $10.5\pm 2.1$\\
\end{tabular}
\caption{Total neutrino fluxes with corresponding uncertainties. The maximum neutrino energy, $E_{\nu}^{\rm{max}}$, and maximum recoil energy of a Xenon target, $E_{r_{\rm{Xe}}}^{\rm{max}}$, are also shown.\label{tab:neutrino}}
\end{ruledtabular} 
\end{table}

\section{Neutrino floor}\label{sec:nufloor}
\begin{figure}[t]
\begin{center}
\includegraphics[trim = 0mm 0 0mm 0mm, clip, width=0.49\textwidth,angle=0]{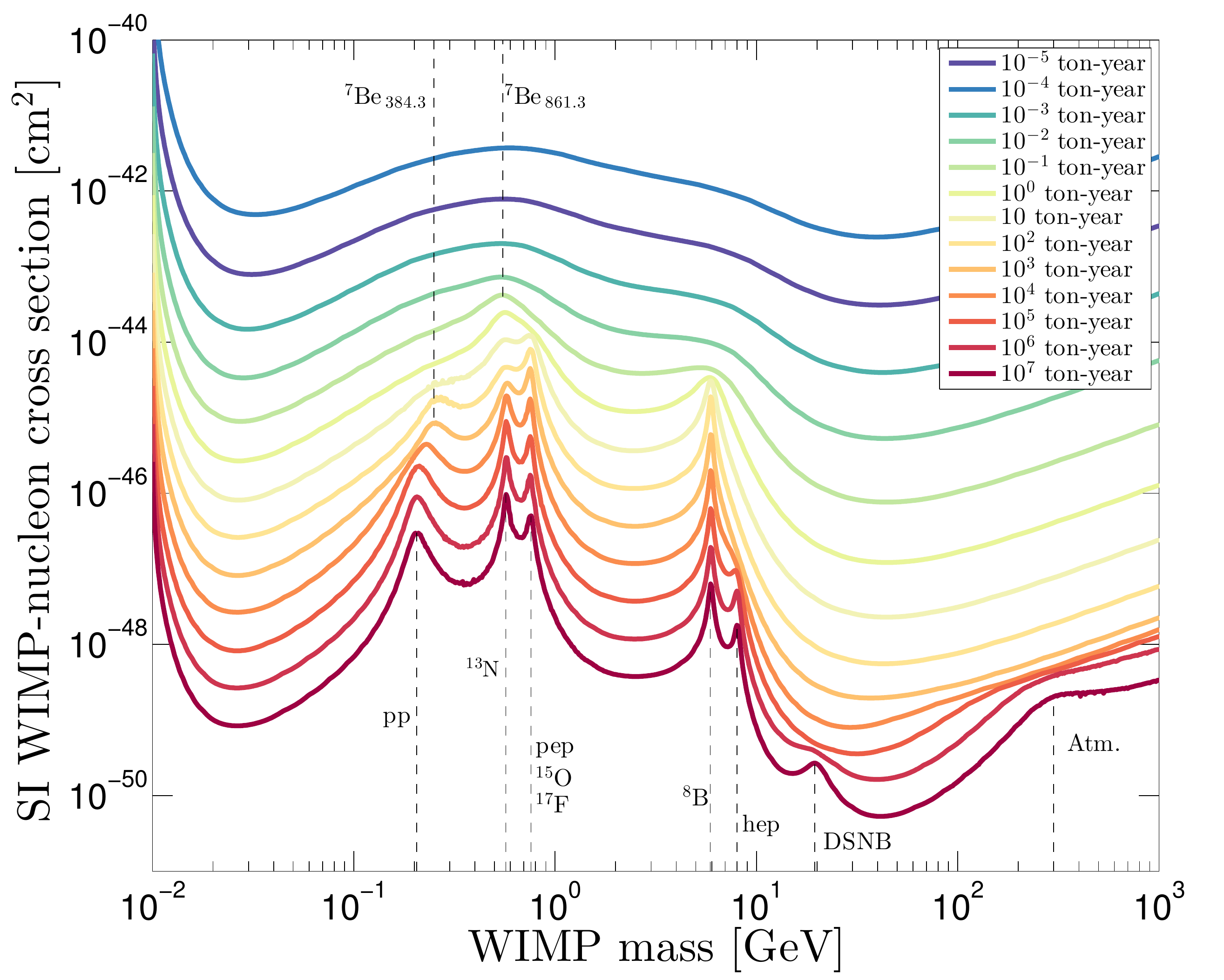}
\caption{Full dependence of the spin-independent neutrino floor for a Xenon target as a function of WIMP mass and detector exposure, showing the contribution from all sources of neutrino. The neutrino floor has peaks at WIMP masses where the Xenon scattering rate for WIMPs and a certain neutrino source overlap.}
\label{fig:EL_vs_mass_allNeutrinos}
\end{center}
\end{figure} 
In this work we will adopt a fixed mock experimental setup with a Xenon target and a 3 eV energy threshold. This is a slightly unrealistic expectation for the sort of threshold possible with a dual phase Xenon detector even beyond the next generation of experiments. We will explore more reasonable expectations for realistic direct detection experiments in Sec.~\ref{sec:future} but initially we make this choice of mock detector setup for a number of reasons. Firstly by using Xenon it allows us to make direct comparisons with previous work on the neutrino floor e.g., Refs.~\cite{neutrinoBillard,Monroe:2007xp,neutrinoRuppin,Grothaus:2014hja,O'Hare:2015mda} whilst also providing us with the simplicity of a single target nucleus. Using a very low threshold also allows us to capture the low WIMP mass neutrino floor without having to change the target nucleus or use targets with multiple different nuclei. Using a constant target and threshold then allows us to isolate the effects due to the dependence on the astrophysical input. 

In this study we will adopt a binned likelihood with $N_\textrm{bins}=100$ to allow us to efficiently extend our analysis to large numbers of neutrino events. The likelihood is written as the product of the Poisson probability distribution function ($\mathscr{P}$) for each bin, multipled by individual likelihood functions parameterising the uncertainties on each neutrino flux normalisation and each astrophysical parameter,
\begin{align}\label{eq:likelihood}
 \mathscr{L}(m_\chi,\sigma_{\chi-n},\boldsymbol{\Phi},\boldsymbol{\Theta}) =& \prod_{i=1}^{N_\textrm{bins}} \mathscr{P} \left(N_\textrm{obs}^i \bigg| N^i_\chi + \sum_{j=1}^{n_\nu} N^{i}_\nu(\phi^j)\right) \nonumber \\ 
& \times \prod_{j=1}^{n_\nu} \mathcal{L}(\phi^j) \nonumber \\
& \times \prod_{k=1}^{n_\theta} \mathcal{L}(\theta^k) \, .
\end{align}
Where $\boldsymbol{\Phi} = \{ \phi^1, ..., \phi^{n_\nu} \}$ are the neutrino fluxes for each of the $n_\nu$ neutrino types and $\boldsymbol{\Theta} = \{\theta^1, ..., \theta^{n_\theta} \}$ contains the $n_\theta$ astrophysical uncertainties under consideration which will vary depending on the choice of velocity distribution for example the standard halo model: $\boldsymbol{\Theta}_\textrm{SHM} = \{ v_0,\vesc,\rho_0 \}$. The functions $\mathcal{L}(\phi^j)$ are the Gaussian parameterisations for each neutrino flux (see Table~\ref{tab:neutrino}) and similarly the likelihood functions $\mathcal{L}(\theta^k)$ parametrise the systematic uncertainty on each astrophysical parameter. Inside the Poisson function we have $N^i_\textrm{obs}$ the number of events observed in bin $i$, as well as $N_\chi^i$ the expected number of WIMP events in the $i$th bin,
\begin{equation}\label{eq:nchi}
N_\chi^i(m_\chi,\sigma_{\chi-n},\boldsymbol{\Theta}) =\mathcal{E}\int_{E_{i}}^{E_{i+1}} \frac{\textrm{d} R_\chi}{\textrm{d}E_r} \textrm{d}E_r \, ,
\end{equation}
and $N_\nu^i(\phi^j)$ is the expected number of neutrino events from the $j$th neutrino species in the $i$th bin,
\begin{equation}
N^{i}_\nu (\phi^j) = \mathcal{E}\int_{E_{i}}^{E_{i+1}} \frac{\textrm{d} R_\nu}{\textrm{d}E_r}(\phi^j) \textrm{d}E_r \, ,
\end{equation}
where $\mathcal{E}$ is the exposure of the experiment which we will quote in units of ton-year.

Limits placed on the WIMP mass-cross section parameter space by a given experiment can be calculated using various statistical methods. The approach taken by Refs.~\cite{neutrinoBillard,neutrinoRuppin,O'Hare:2015mda} uses a profile likelihood ratio test statistic to calculate so-called ``discovery limits'' which are defined as the minimum cross section for which a $3\sigma$ discovery of a WIMP is possible in 90\% of hypothetical experiments. The profile likelihood ratio comprises a hypothesis test between the null hypothesis $H_0$ (neutrino only) and the alternative hypothesis $H_1$ which includes both neutrinos and a WIMP signal whilst incorporating systematic uncertainties, in this case on the flux of each neutrino component $\boldsymbol{\Phi}$ and the astrophysical parameters $\boldsymbol{\Theta}$. We can then test the background only hypothesis, $H_0$, on a simulated dataset by attempting to reject it using the likelihood ratio,
\begin{equation}
\lambda(0) = \frac{\mathscr{L}(\sigma_{\chi-n} = 0,\hat{\hat{\boldsymbol{\Phi}}},\hat{\hat{\boldsymbol{\Theta}}})}{\mathscr{L}(\hat{\sigma}_{\chi-n},\hat{\boldsymbol{\Phi}},\hat{\boldsymbol{\Theta}})},
\end{equation}
where ${\hat{\boldsymbol{\Phi}}}$, ${\hat{\boldsymbol{\Theta}}}$ and $\hat{\sigma}_{\chi-n}$ denote the values of ${\boldsymbol{\Phi}}$, ${\boldsymbol{\Theta}}$ and $\sigma_{\chi-n}$ that maximise the unconditional $\mathscr{L}$ and  
$\hat{\hat{\boldsymbol{\Phi}}}$ and $\hat{\hat{\boldsymbol{\Theta}}}$ denote the values of ${\boldsymbol{\Phi}}$ and ${\boldsymbol{\Theta}}$ that maximise $\mathscr{L}$ under the condition $\sigma_{\chi-n} = 0$, i.e., we are profiling over the nuisance parameters ${\boldsymbol{\Phi}}$ and ${\boldsymbol{\Theta}}$. Note that the test is conducted at fixed WIMP mass and then repeated over a range of input masses. As introduced in Ref.~\cite{Cowan:2010js}, the test statistic $q_0$ is then defined as,
\begin{equation}
q_0 = \left\{
\begin{array}{rrll}
\rm & -2\ln\lambda(0)	&	\ \hat{\sigma}_{\chi-n} > 0 \,, \\
\rm & 0  		& 	\ \hat{\sigma}_{\chi-n} < 0 \, . 
\end{array}\right.
\end{equation}
If a large value of $q_0$ is calculated then this implies that the alternative hypothesis gives a better fit to the data and the existence of a WIMP signal is preferred. The $p$-value, $p_0$, of a particular experiment is the probability of finding a value of the test statistic larger than or equal to the observed value, $q_0^{\rm obs}$, if the null hypothesis is correct,
\begin{equation}
p_0 = \int_{q_0^{\rm obs}}^{\infty} f(q_0|H_0) \, {\rm d} q_0 \, , 
\end{equation}
where $f(q_0|H_0)$ is the probability distribution function of the test statistic under the background only hypothesis. From Wilk's theorem~\cite{Cowan:2010js}, $q_0$ asymptotically follows a $\chi^2$ distribution with one degree of freedom and therefore the significance, $Z$, in units of Gaussian standard deviation ($\sigma$) is simply given by $Z = \sqrt{q^{\rm obs}_0}$. The discovery limit for a particular input WIMP mass is then found by finding the smallest input cross section for which 90\% of simulated experiments have $Z \geq 3$.

When discussing WIMP and neutrino detection the term ``neutrino floor'' is loosely applied to the region of WIMP parameter space for which neutrinos become a problematic background. But in fact how this limit evolves as a function of detector exposure is slightly more complex, as we will now briefly discuss. In the case of an experiment which only records event number, for example a bubble chamber, the minimum discoverable cross section at a given fixed WIMP mass plateaus as the experiment detects an increasing number of events and for large numbers of events there can be no background subtraction. This is due to the systematic uncertainty on the neutrino flux; small cross sections which induce an excess in the number of events that is smaller than that due to fluctuations around the expected neutrino flux cannot be attributed to a WIMP with the required significance. In a hypothetical case in which the expected neutrino event rate was known with perfect certainty there would be no limit to how small a cross section the experiment could discover and as the number of background events increased the discovery limit would keep decreasing according to well understood Poissonian background subtraction. However with a systematic uncertainty the neutrino background gives rise to a ``floor'' - a limit that divides the $m_\chi$-$\sigma_{\chi-n}$ space into WIMPs which are accessible to the experiment and those which are indistinguishable from neutrinos. This problem persists even with the addition recoil energy information because of the similarity in the spectra of WIMP and neutrino induced recoils~\cite{neutrinoRuppin}. However the slight differences in the tails of the neutrino and WIMP event rates allows the two spectra to be distinguished once a sufficient number of events has been detected, usually around the order of $\mathcal{O}(1000)$ events (though the precise number depends on the size of the neutrino flux uncertainty).

Figure~\ref{fig:EL_vs_mass_allNeutrinos} shows the full evolution of the discovery limit for a Xenon experiment with an extremely low threshold (0.01 eV) to capture the neutrino floor down to $pp$ neutrino energies for completeness. This is a similar plot to a result of Ref.~\cite{neutrinoRuppin} but here we use updated values for the neutrino fluxes and uncertainties and extend to a lower threshold and to larger exposures. The floor moves to lower cross sections as the exposure is increased as one would expect, however it aquires peaks where the WIMP recoil spectrum is mimicked by a given neutrino component. The mass at which a peak appears is dependent on the recoil energy range of the neutrino type. The cross section of the peak and how long the peak remains as exposure is increased depends on the uncertainty on the neutrino flux as well as how well the WIMP recoil event rate is mimicked by the neutrino type~\cite{neutrinoRuppin}. With a smaller uncertainty it takes fewer WIMP events to distinguish them from neutrinos. The most prominent contribution to the neutrino floor is due to $^8$B neutrinos which cause the floor to appear at 6 GeV. There are also contributions from $hep$, atmospheric and DSNB neutrinos at higher WIMP masses and at the low WIMP mass end (below 1 GeV) there is a cluster of peaks due to the lower energy Solar neutrinos: $pp$, $pep$, $^7$Be, $^{15}$O, $^{13}$N and $^{14}$F.

The neutrinos that are most pertinent for direct detection searches are $^8$B. These are expected to be the first type of neutrino to be detected through coherent neutrino-nucleus scattering as they induce recoil energies within the scope of the next generation of detector and do not require unfeasibly large exposures to observe a significant number of events. However when studying the effects of the neutrino background on direct detection one must make a choice between an unrealistically low threshold in order to observe the low energy Solar neutrinos, or an unrealistically large exposure in order to observe atmospheric and DSNB neutrinos. In this work as we are interested in the role played by the astrophysics dependence of the WIMP signal we will make the former choice. This is because light WIMPs are the more phenomenologically interesting region in this context - they probe the tail of the speed distribution and this is typically where there is the most senstivity to the values of astrophysical parameters. This is also coupled with the fact that advances in technology are more likely to bring about lower threshold detectors than allow exposures in excess of $10^6$ ton-years to be achieved (which are required to observe the neutrino floor due to DSNB and atmospheric neutrinos). For instance, a recent work by Mirabolfathi~\etal~\cite{Mirabolfathi:2015pha} outlined how with current advances in technology, ultra-low thresholds down to $\sim 10$ eV may be achievable in cryogenic detectors with excellent energy resolution.

\section{Astrophysical uncertainties}\label{sec:astro}
\begin{figure}[t]
\begin{center}
\includegraphics[trim = 0mm 0 0mm 0mm, clip, width=0.49\textwidth,angle=0]{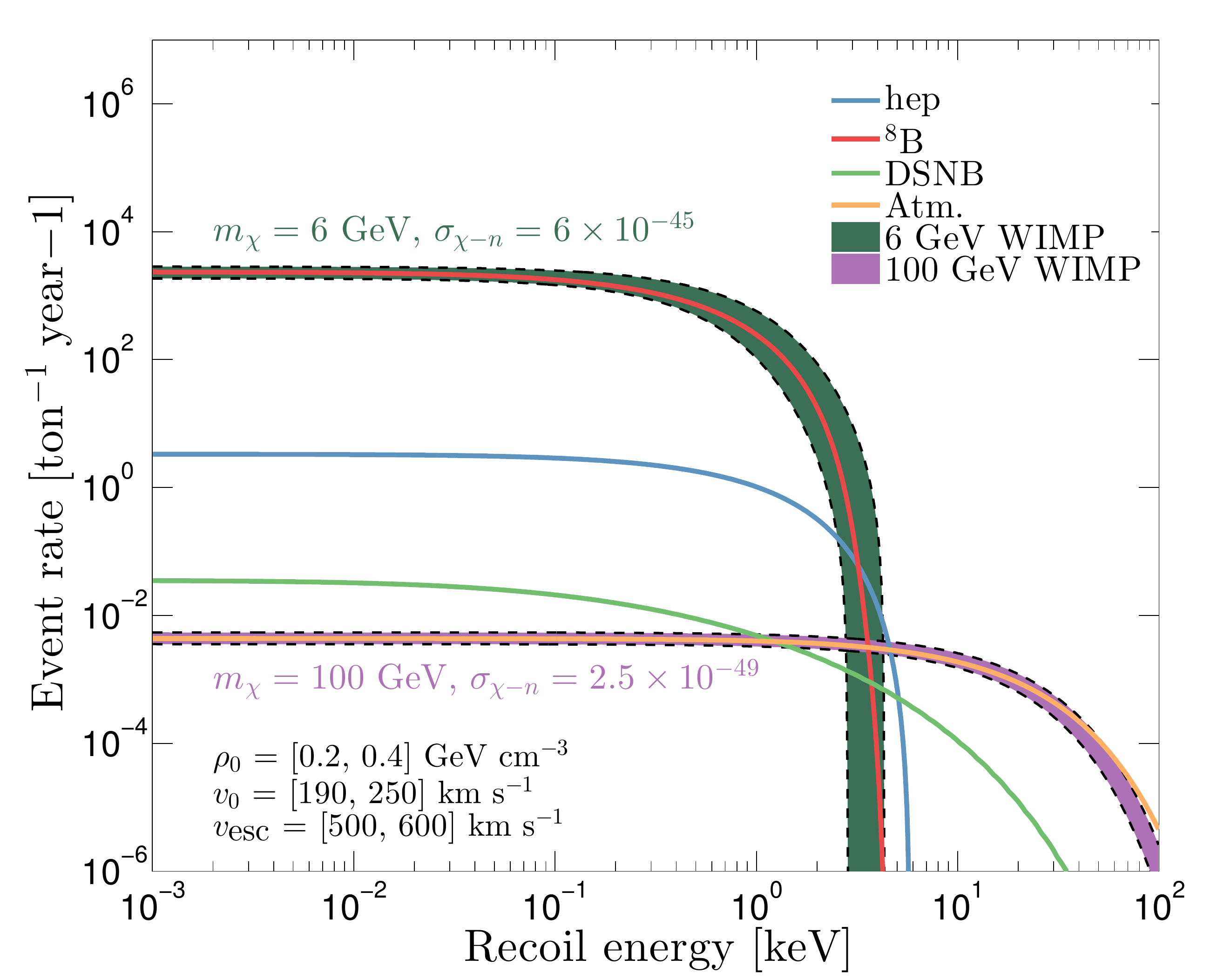}
\caption{Spin-independent Xenon elastic scattering rates for 6 and 100 GeV WIMPs and 4 neutrino sources ($hep$, $^8$B, DSNB and atmospheric neutrinos). The dark green and orange shaded regions respectively refer to the range of scattering rates for 6 and 100 GeV WIMPs with standard halo model parameters taking values between $\rho_0 = [0.2,0.4]$~GeV~cm$^{-3}$, $v_0 = [190,250]$~km~s$^{-1}$ and $\vesc = [500,600]$~km~s$^{-1}$.}
\label{fig:EventRates_with_uncertainties}
\end{center}
\end{figure} 
The simplest approximation of a dark matter halo is known as the standard halo model (SHM): an isotropic and isothermal sphere of dark matter with a $1/r^2$ density profile in which the Milky Way stellar disk is embedded. The velocity distribution of dark matter yielded by such a model has a Maxwell-Boltzmann form and is usually truncated at the escape speed of the Galaxy,
\begin{equation}\label{eq:shm}
f({\bf v}) = \left\{
\begin{array}{llrr}
\frac{1}{N} \,e^{-v^2/v_0^2} &	\ \text{if $|{\bf v}|<\vesc$\,,}  \\
0  		& 	\ \text{if $|{\bf v}|\geq \vesc$\,.}
\end{array}\right.
\end{equation}
Where $\vesc$ is the escape speed, $v_0$ is the circular rotation speed of the Galaxy and $N$ is a normalisation constant found by imposing $\int f(\textbf{v}) \, \textrm{d}^3 v = 1$. Figure~\ref{fig:EventRates_with_uncertainties} shows the energy dependence of the nuclear recoil event rate over a range of input values for the three free parameters of this model: local density $\rho_0$, circular rotation speed $v_0$ and escape velocity $\vesc$. As mentioned in Sec~\ref{sec:nufloor} we see that light WIMPs have a greater sensitivity to changes in the astrophysical input than heavier WIMPs and that the most visible change is around the tail of the recoil distribution (around 1 keV for a 6 GeV WIMP for example). We will first consider the effect of each parameter of the SHM individually in Secs.~\ref{sec:escapevelocity}-~\ref{sec:localdensity}, as well as the assumption for the speed distribution in Sec.~\ref{sec:velocitydistribution}, before combining all sources of uncertainty in Sec.~\ref{sec:astrounc}.

\subsection{Escape velocity}\label{sec:escapevelocity}
\begin{figure}[t]
\begin{center}
\includegraphics[trim = 0mm 0 0mm 0mm, clip, width=0.49\textwidth,angle=0]{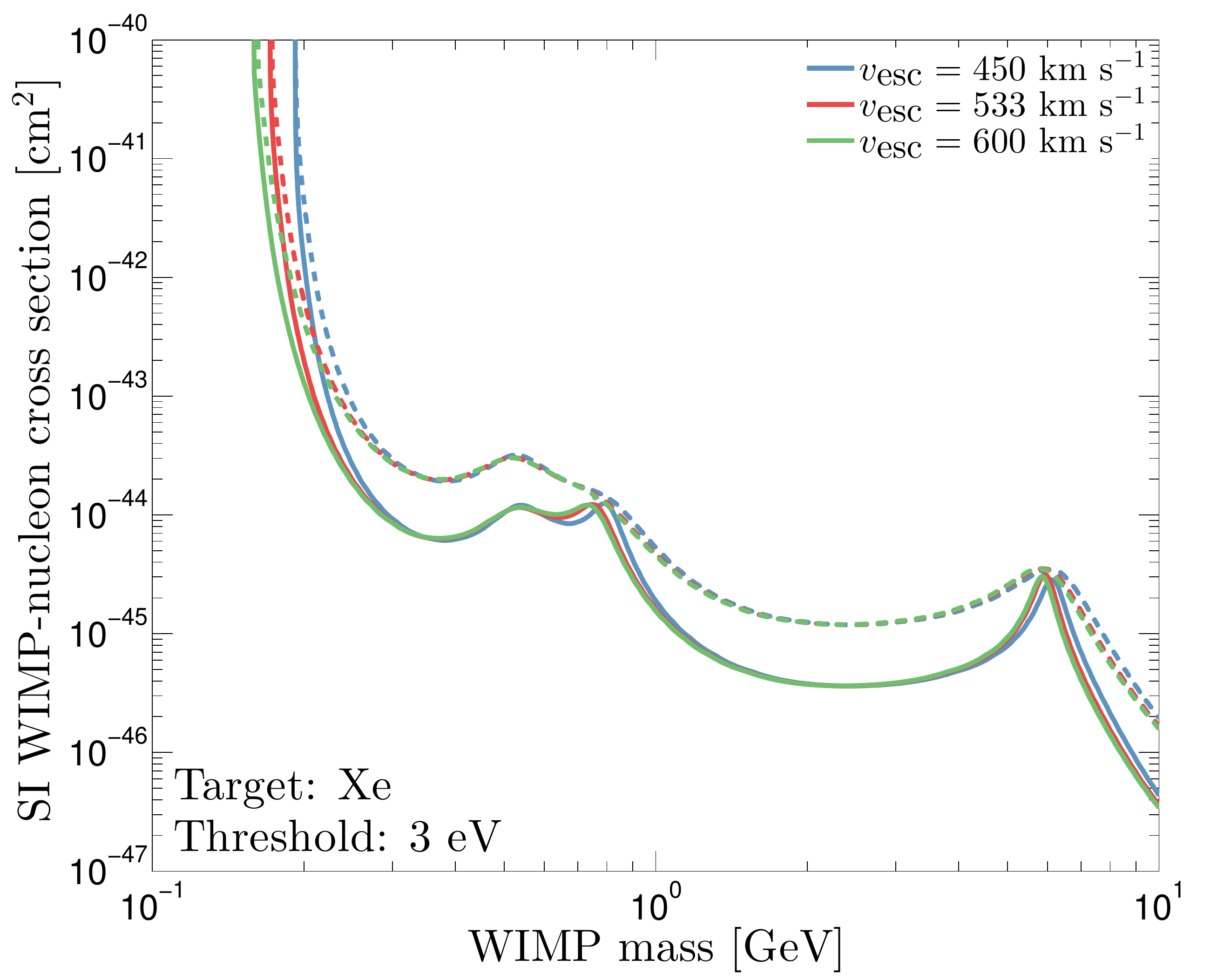}
\caption{Spin-independent discovery limit for a Xenon experiment with different values of input escape velocity. The dashed lines are for a 1 ton-year exposure and the solid lines for a 10 ton-year exposure. The blue, red and green colours correspond to input escape velocities of 450, 533 and 600 km s$^{-1}$ respectively.}
\label{fig:DL_vesc}
\end{center}
\end{figure} 

The escape velocity is the maximum speed a dark matter particle can have whilst still being considered gravitationally bound to the Milky Way. It in principle sets the maximum WIMP speed that can be detected on Earth. The escape velocity can be measured directly by finding high velocity stars in the Milky Way to attempt to map the tail of the global Galactic speed distribution~\cite{Leonard1990}. Alternatively it can be inferred by calculating the gravitational potential of the Galaxy using astronomical data. The most noteworthy estimates to date have been made using data from the RAVE survey~\cite{Steinmetz:2006qt} first released in 2006. An estimate of $\vesc=544^{+65}_{-46}$km s$^{-1}$~\cite{Smith:2006ym} was made using the first release of this data and is commonly used to derive many direct detection exclusion limits, but the most recent estimate from 2014 based on the fourth release of RAVE data finds $\vesc=533^{+54}_{-41}$ km s$^{-1}$~\cite{Piffl:2013mla}.

Since the escape velocity can only control the tail of the recoil distribution and because the speed distribution is very small at its tail, the effect of changing the speed at which it is truncated only has a small effect on the overall shape of the recoil energy spectrum. However the escape velocity plays an important role in dictating the smallest WIMP mass detectable by an experiment. Most depictions of discovery limits show a sharp increase at low WIMP masses when the maximum recoil energy (set by the maximum WIMP speed) falls below the threshold of the experiment. As such we expect changes in the escape velocity to show up most prominently in this sharply increasing low mass region. Figure~\ref{fig:DL_vesc} shows the neutrino floor for 3 values of the escape velocity. We can see here that changing the escape velocity has a very marginal effect on the overall shape of the discovery limits. The most noticeable effect is around 0.2 GeV where the discovery limit sharply increases due to the maximum energy recoils falling below 3 eV. For smaller values of escape velocity this sharp increase in the discovery limit appears at a larger WIMP mass. However above 0.2 GeV and between 1 and 5 GeV the discovery limits for different values of $\vesc$ are indistinguishable. This result agrees with the findings of McCabe~\cite{McCabe:2010zh}, however there are now small differences around 6 GeV and 0.8 GeV when the tails of the recoil energy distributions become important in discriminating between the WIMP and neutrino signals. However these differences are extremely minor. 

It should be noted that the values of escape velocity chosen in Fig.~\ref{fig:DL_vesc} cover a wider range than the expected uncertainty in the central value of 533 km s$^{-1}$ so given these results we deduce that including the uncertainty in $\vesc$ will only have a small effect. When in subsequent results (Sec~\ref{sec:astrounc}) we include uncertainties in the statistical analysis, we will use the central RAVE value and a range of uncertainties up to the values quoted in the literature.

\subsection{Solar velocity}\label{sec:solarvelocity}
\begin{figure}[t]
\begin{center}
\includegraphics[trim = 0mm 0 0mm 0mm, clip, width=0.49\textwidth,angle=0]{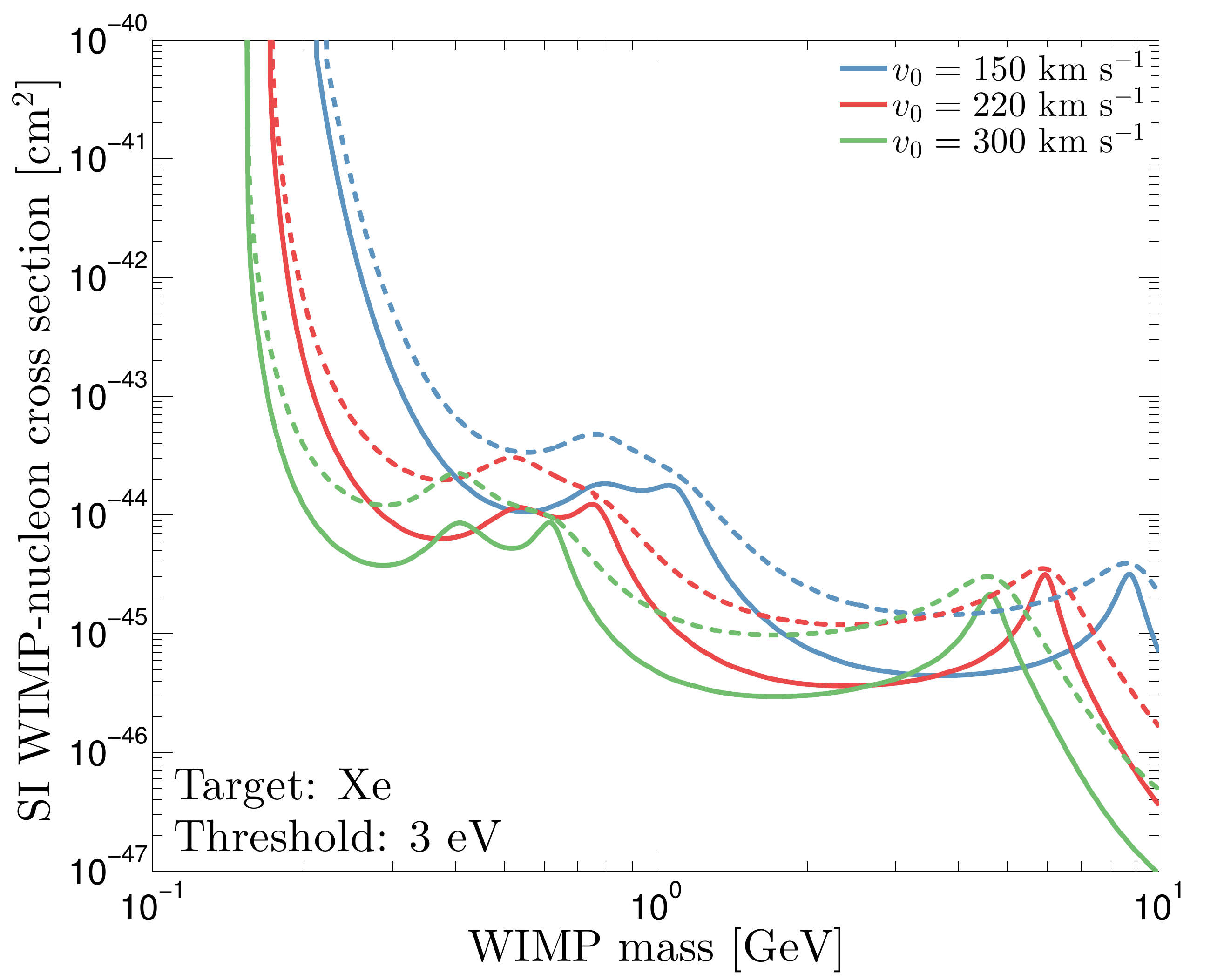}
\caption{Spin-independent discovery limits for a Xenon experiment with different values of input circular velocity $v_0$. The dashed lines are for a 1 ton-year exposure and the solid lines for a 10 ton-year exposure. The blue, red and green colours correspond to an input $v_0$ of 150, 220 and 300 km s$^{-1}$ respectively.}
\label{fig:DL_v0}
\end{center}
\end{figure} 
The velocity distribution is observed through a Galilean boost into the laboratory frame by the velocity with which we are moving with respect to the halo, $\textbf{v}_\textrm{lab}$. This velocity is the sum of 4 components: the bulk velocity of the Milky Way local standard of rest (LSR) $\textbf{v}_0$, the peculiar velocity of the Sun with respect to the LSR $\textbf{v}_\odot$, the velocity of the Earth with respect to the Sun $\textbf{v}_\textrm{rev}$, and the rotation of the Earth $\textbf{v}_\textrm{rot}$. The latter two velocities are responsible for the annual~\cite{annual} and diurnal~\cite{daily} modulations in the event rate respectively and are known theoretically with effectively perfect precision. The peculiar velocity is believed to possess a reasonably small uncertainty. A value commonly used from Schoenrich \etal~\cite{Schoenrich:2009bx} gives $\textbf{v}_\odot = (11.1,12.24,7.25)$~km~s$^{-1}$ in Galactic co-ordinates with roughly 1~km~s$^{-1}$ sized systematic errors. In this section we will ignore these contributions to the laboratory velocity, however when we incorporate the time dependencies of the WIMP and Solar neutrino event rates in Sec.~\ref{sec:time} the Earth revolution and rotation velocities will be included. 

The largest source of uncertainty in the Laboratory velcocity comes from the Sun's circular speed. It is also the largest contribution to $\textbf{v}_\textrm{lab}$ at roughly $v_0\sim220$~km~s$^{-1}$ which is the fiducial value usually used~\cite{Kerr:1986hz}. The circular speed has been measured in various ways, for instance measurements of the proper motions of objects such as nearby stars or Sgr A* located at the Galactic centre can be used to constrain the quantity $(v_0+V_\odot)/R_\odot$ where $V_\odot$ is the second component of $\textbf{v}_\odot$ and $R_\odot$ is the Solar Galactic radius. Given independent constraints on the Sun's peculiar velocity and radius one can combine measurements to arrive at a constraint on $v_0$. However, as noted by Lavalle and Magni~\cite{Lavalle:2014rsa} because these estimates depend upon the prior assumptions made about other parameters, combining measurements of, for instance $R_\odot$ and $V_\odot$ from different scources may lead to spurious resulting values and underestimated errors. Ref.~\cite{Reid:2014boa} however contains an estimate for $v_0$ of $243 \pm 6$~km~s$^{-1}$ which makes use of the same priors on Solar motion as the study based on the RAVE data (Ref.~\cite{Schoenrich:2009bx}), meaning its value is consistent with $\vesc=533$~km~s$^{-1}$.

Given the discrepencies between astronomically observed values for $v_0$, both with each other and with the fiducial value of 220~km~s$^{-1}$, we will be pessimistic about our chosen uncertainty on $v_0$. Figure~\ref{fig:DL_v0} shows the neutrino floor for a range of values of $v_0$, keeping other parameters constant. Comparing with Fig.~\ref{fig:DL_vesc} we can see the effect of $v_0$ is much more noticeable. Whereas $\vesc$ affects only the tail of the recoil energy distribution, $v_0$ affects the entirety. As explained in Sec.~\ref{sec:nufloor}, the shape of the neutrino floor is determined by the WIMP masses which scatter into energies that overlap with each neutrino component. For smaller values of $v_0$ larger WIMP masses are needed to produce recoils which are mimicked by the same neutrino type, it is understandable then that for smaller $v_0$, the neutrino floor is shifted to higher WIMP masses. Additionally with smaller $v_0$ more of the recoil distribution falls below the threshold and this reduction in the number of events causes a shift in the floor to larger cross sections i.e., the experiment is less powerful at a given WIMP mass.
 
\subsection{Local density}\label{sec:localdensity}
The local density of WIMPs $\rho_0$, is most often taken to be its fiducial value of 0.3~GeV~cm$^{-3}$. This is mostly because it appears as a multiplicative factor in the WIMP event rate and is as a result degenerate with the scattering cross section. Moreover, calculations of the local density have been historically variable. Recent work by Ref.~\cite{Bienayme:2014kva} find a value of $0.542\pm0.042$ GeV cm$^{-3}$ using a host of red clump stars from RAVE observations, whereas Ref.~\cite{Lavalle:2014rsa} find values between 0.42 and 0.08 depending on the choice of prior on $v_0$. For us however the effect of changing local density is straightforward; a larger value of $\rho_0$ simply shifts the floor to smaller cross sections by the same factor.

\subsection{Speed distribution}\label{sec:velocitydistribution}
Most direct detection analyses use the SHM Maxwellian speed distribution both for simplicity as well as to establish a baseline on which to compare different experiments (given that there is no fully agreed upon alternative). Nevertheless it is well known from simulations that the SHM is not a good description of a Milky Way-like halo. There have been numerous attempts to find empirical fitting functions to better capture the phase space structure found in N-body and hydrodynamic simulations~\cite{Mao:2012hf,Lisanti:2010qx,Ling:2009eh,Bozorgnia:2016ogo,Kelso:2016qqj,Sloane:2016kyi} as well as parameterisations that decompose the speed or velocity distribution in an astrophysics independent way~\cite{Kavanagh:2013eya,Lee:2014cpa}. Some studies of data from hydrodynamic simulations suggest that the standard halo model is a satisfactory approximation to the Milky Way once baryons are taken into account (e.g., Ref.~\cite{Kelso:2016qqj}), however others such as Sloane~\etal~\cite{Sloane:2016kyi} claim that the SHM overpredicts the amount of dark matter in the tail and hence gives overly optimistic discovery limits. To address these concerns, and because when discriminating between WIMPs and neutrinos the high speed tail of the distribution is especially important, we show discovery limits for a range of different models. Here we describe three examples that we use to serve as a demonstration of the effect of the input speed distribution on the neutrino floor. These cover a reasonable range of possible parameterisations with the exception of speed distributions that contain additional features such as tidal streams which we leave for future work.

Halos with double power law density profiles, such as the NFW profile, can have their high velocity dependence better reproduced if a distribution is chosen of the form~\cite{Stadel:2008pn},
\begin{equation}
f_{\rm DPL}({\bf v}) = \left\{
\begin{array}{llrr}
 \frac{1}{N}\left[\exp{\left(-\frac{\vesc^2-v^2}{kv_0^2}\right)}-1\right]^k &	\ \text{if $|{\bf v}|<\vesc$\,,}  \\
0   		& 	\ \text{if $|{\bf v}|\geq \vesc$\,.}
\end{array}\right.
\end{equation}
This model is a modification of the SHM (the form of Eq.~\ref{eq:shm} is recovered when setting $k=1$). Results from N-body simulations suggest $k$ to be in the range $1.5<k<3.5$~\cite{Busha:2004uk,Lisanti:2010qx}.

In Ref.~\cite{Ling:2009eh} it was found that the Tsallis model produced a better fit to simulations which included baryons. It involves a speed distribution of the form, 
\begin{equation}
f_{\rm Tsallis}({\bf v}) = \left\{
\begin{array}{llrr}
 \frac{1}{N}\left[1-(1-q)\frac{v^2}{v_0^2}\right]^{1/(1-q)} &	\ \text{if $|{\bf v}|<\vesc$\,,}  \\
 0  		& 	\ \text{if $|{\bf v}|\geq \vesc$\,.}
\end{array}\right.
\end{equation}
with best fit parameters of $q = 0.773$, $v_0 = 267.2$~km~s$^{-1}$ and $\vesc = 560.8$~km~s$^{-1}$.

The final speed distribution we consider is one introduced by Mao \etal~\cite{Mao:2012hf,Mao:2013nda}. Which was found to improve the fit in simulations. It takes a form characterised by an index $p$,
\begin{equation}
f_{\rm Mao}({\bf v}) = \left\{
\begin{array}{llrr}
 \frac{1}{N}\,e^{-v/v_0}\left(\vesc^2-v^2\right)^p &	\ \text{if $|{\bf v}|<\vesc$\,,}  \\
0   		& 	\ \text{if $|{\bf v}|\geq \vesc$\,.}
\end{array}\right.
\end{equation}
Where results from the Rhapsody and Bolshoi simulations give $p$ in the range $0<p<3$.

\begin{figure}[t]
\begin{center}
\includegraphics[trim = 0mm 0 0mm 0mm, clip, width=0.49\textwidth,angle=0]{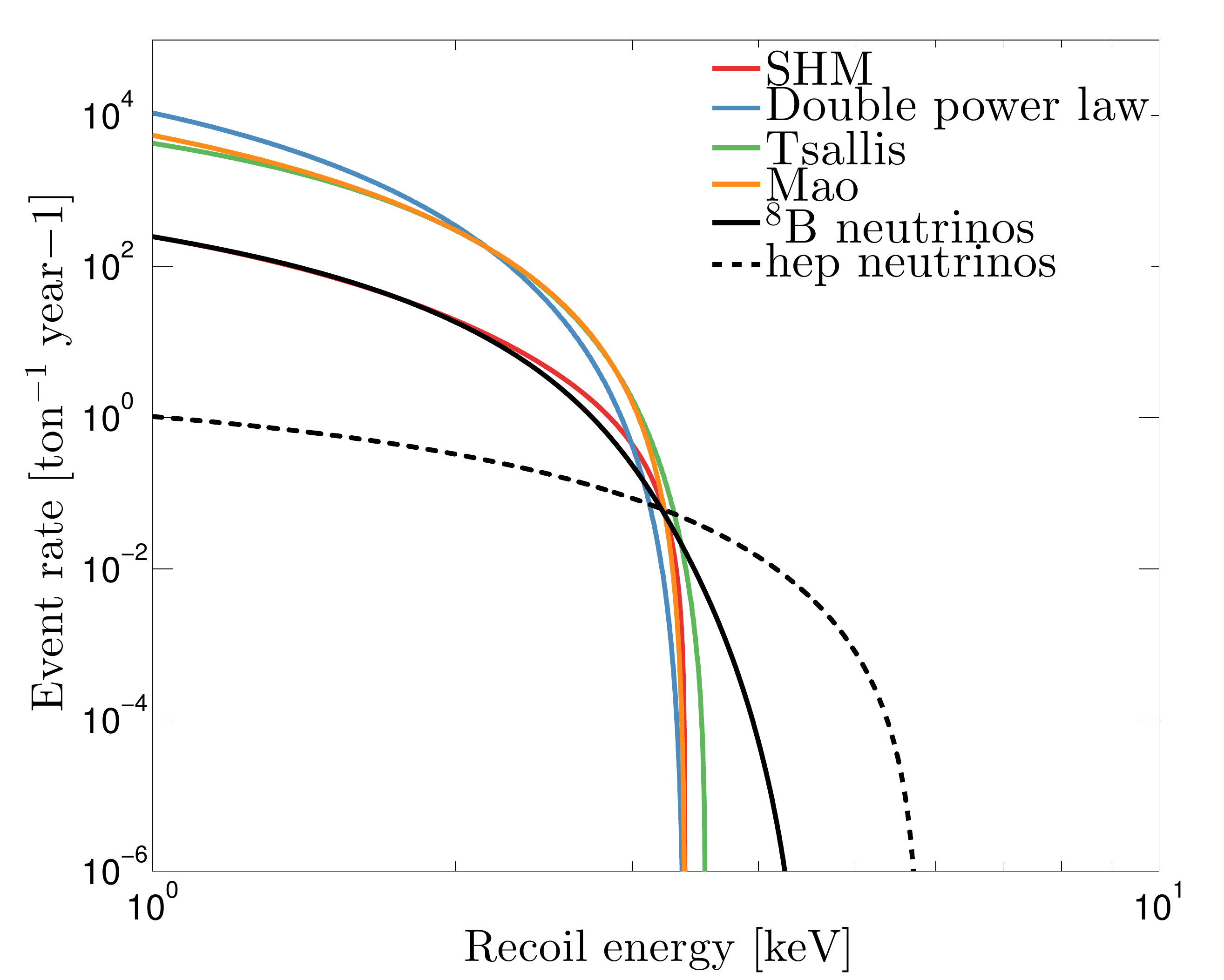}
\caption{Event rate as a function of energy for Xenon recoils and a 6 GeV WIMP with a spin-independent cross section of $\sigma_{\chi-n} = 5\times10^{-45}$~cm$^2$. Shown are the event rates computed for 4 speed distributions as described in the text: the standard halo model, double power law, Tsallis, and Mao models (red, blue, green and orange curves respectively). Also shown is the recoil energy distributions from $^8$B and $hep$ neutrinos in the solid and dashed black curves respectively.}
\label{fig:EventRates_speeddists}
\end{center}
\end{figure}
\begin{figure}[t]
\begin{center}
\includegraphics[trim = 0mm 0 0mm 0mm, clip, width=0.49\textwidth,angle=0]{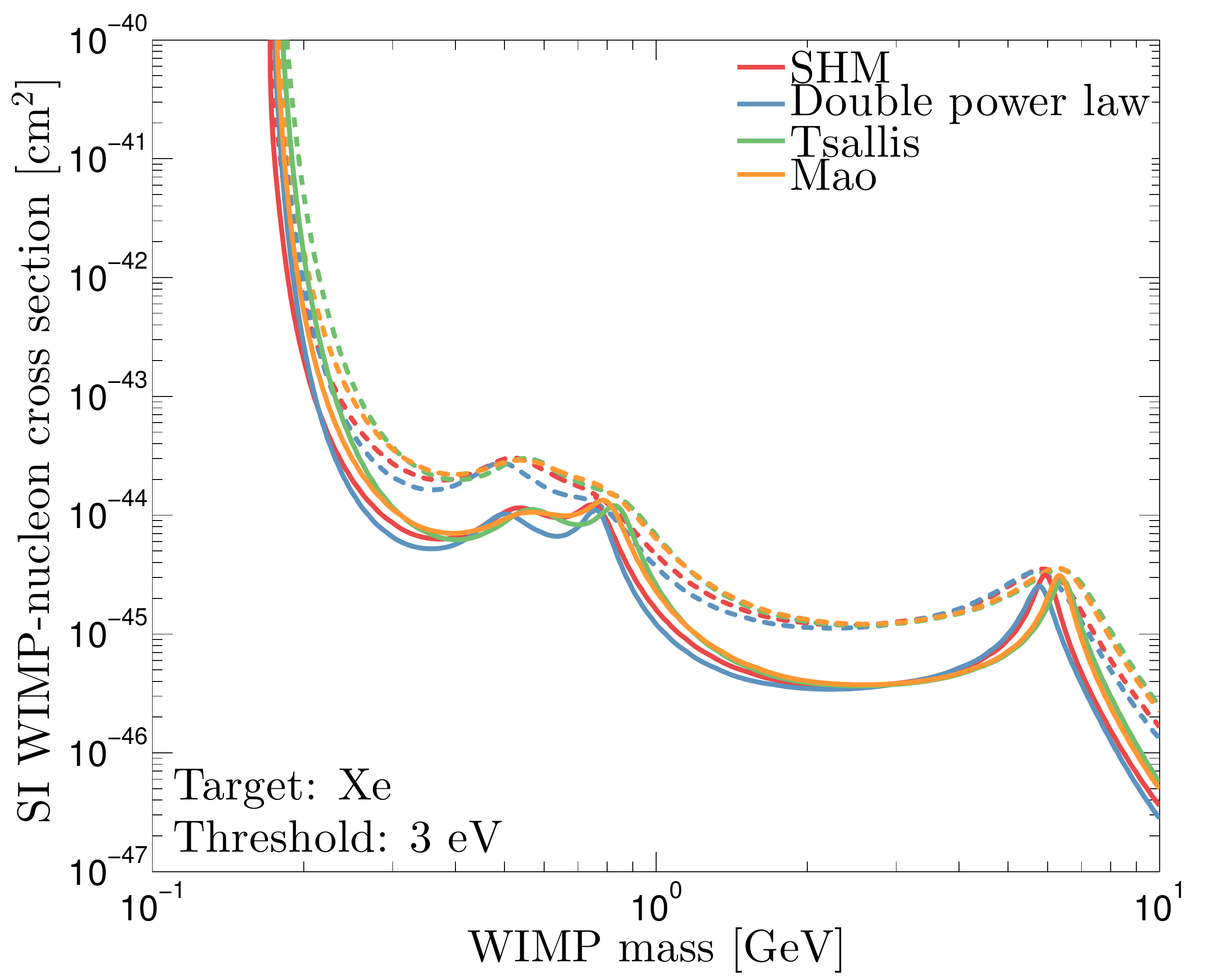}
\caption{Spin-independent discovery limit for a Xenon experiment for different input speed distributions. The dashed lines are for a 1 ton-year exposure and the solid lines for a 10 ton-year exposure. The blue, green and orange colours correspond to the Double Power Law, Tsallis and Mao distributions respectively. The red lines are for the SHM with $v_0 = 220$~km~s$^{-1}$ and $v_\textrm{esc} = 533$~km~s$^{-1}$.}
\label{fig:DL_fv}
\end{center}
\end{figure}

 
Figure~\ref{fig:EventRates_speeddists} shows the Xenon elastic scattering rate as a function of energy for the three alternative speed distributions compared with the SHM result. For the quoted estimates on their respective parameters the 3 alternative speed distributions have roughly similar shapes with most significant differences occuring around the tail of the distribution close to $\vesc$. The neutrino bounds obtained under the assumption of these alternative distributions are shown in Fig.~\ref{fig:DL_fv}. When the underlying speed distribution is changed the position of the floor shifts only very slightly. Shifting to slightly higher WIMP mass for the double power law and Mao models and to slightly smaller WIMP masses for the Tsallis model. Hence for this reason, and in the interest of efficiency, from here we will neglect the dependence on the speed distribution and focus our attention on reconstructing the parameters of the SHM. 

\subsection{Discovery limits with uncertainties}\label{sec:astrounc}
\begin{figure}[t]
\begin{center}
\includegraphics[trim = 0mm 0 0mm 0mm, clip, width=0.49\textwidth,angle=0]{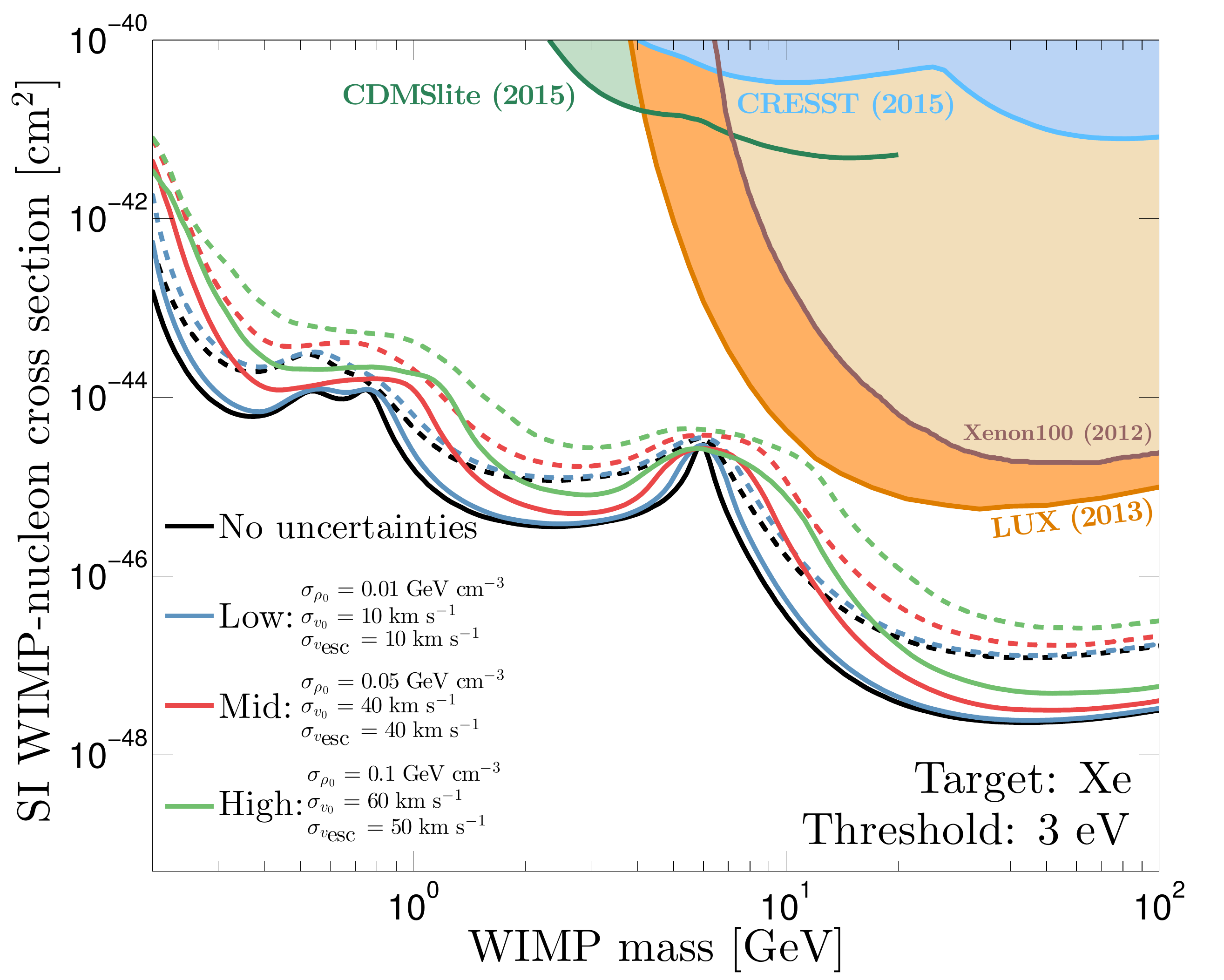}
\caption{Spin-independent discovery limit as a function of WIMP mass calculated with the inclusion of astrophysical uncertainties in the profile likelihood analysis. The dashed lines are for an exposure of 1 ton-year and the solid lines are with an exposure of 10 ton-years. The blue, red and green curves correspond to 3 sets of values of the 1$\sigma$ uncertainty on the parameters $\rho_0$, $v_0$ and $\vesc$ displayed on the Figure and in the text. The size of the uncertainties are labelled from low to high with values indicated. The filled regions are currently excluded by experiments, CRESST~\cite{Angloher:2015ewa}, CDMSlite~\cite{Agnese:2015nto}, Xenon100~\cite{Aprile:2012nq} and LUX~\cite{Akerib:2013tjd}.}
\label{fig:DL_uncertainties}
\end{center}
\end{figure}

Now that we have demonstrated the effect of each parameter individually on the neutrino floor we will unfix the astrophysics parameters in the profile likelihood ratio test and account for their uncertainty with a multiplicative Gaussian parameterisation. Figure~\ref{fig:DL_uncertainties} shows the discovery limits as a function of the width of the Gaussian uncertainty in each parameter. We label the sets of the uncertainty values ``low'', ``mid'' and ``high''. The low values for the 1$\sigma$ uncertainty on $\rho_0$, $v_0$ and $\vesc$ are respectively, 0.01~GeV~cm$^{-3}$, 10~km~s$^{-1}$ and 10~km~s$^{-1}$. The mid values are 0.05~GeV~cm$^{-3}$, 40~km~s$^{-1}$ and 40~km~s$^{-1}$. And for the high values we use 0.1~GeV~cm$^{-3}$, 60~km~s$^{-1}$ and 50~km~s$^{-1}$.

With the values of the astrophysical parameters uncertain, the experiment is less powerful and the discovery limits appear at larger cross sections as more events are needed to make a discovery with the same significance. What we also find is that the peaks that appear in the discovery limit from each neutrino component become broader with the inclusion of uncertainties. We interpret this as being due to the fact that as shown in the previous section for different values of $v_0$ and $v_\textrm{esc}$, the peak in the discovery limit shifts to WIMP masses with recoil energy spectra more closely matching that of the relevent neutrino. The larger the uncertainty on $v_0$ and $v_\textrm{esc}$ the broader the peak becomes. In other words, the greater allowed range of astrophysical parameter values, the wider the range WIMP masses whose recoils overlap with neutrinos. An interesting consequence of this is that the value the floor takes at a WIMP mass of 6 GeV actually {\it decreases} as the uncertainty in $v_0$ is increased, however this is compensated by a large increase for masses above and below 6 GeV.

We have shown here that it is important that the astrophysical input to calculations of discovery limits must be well understood if one wishes to interpret how neutrinos play a role in the discoverability of certain regions of the WIMP mass-cross section parameter space. Particularly this will be a concern for the next generation of direct detection experiments which are set to make limits that come very close to the limits we have calculated here. In fact as we can see in Fig.~\ref{fig:DL_uncertainties}, the limits we have calculated for the ``high'' values of uncertainty come extremely close to the existing LUX limit just above 10 GeV. Hence we can conclude that unless there are improvements in the knowledge of the astrophysics parameters or the uncertainties on the neutrino flux, the neutrino floor will be encountered by direct detection experiments much sooner than previously thought.

\section{Parameter constraints}\label{sec:parcon}
\begin{figure*}[t]
\begin{center}
\includegraphics[trim = 15mm 10mm 15mm 0mm, clip, width=0.9\textwidth,angle=0]{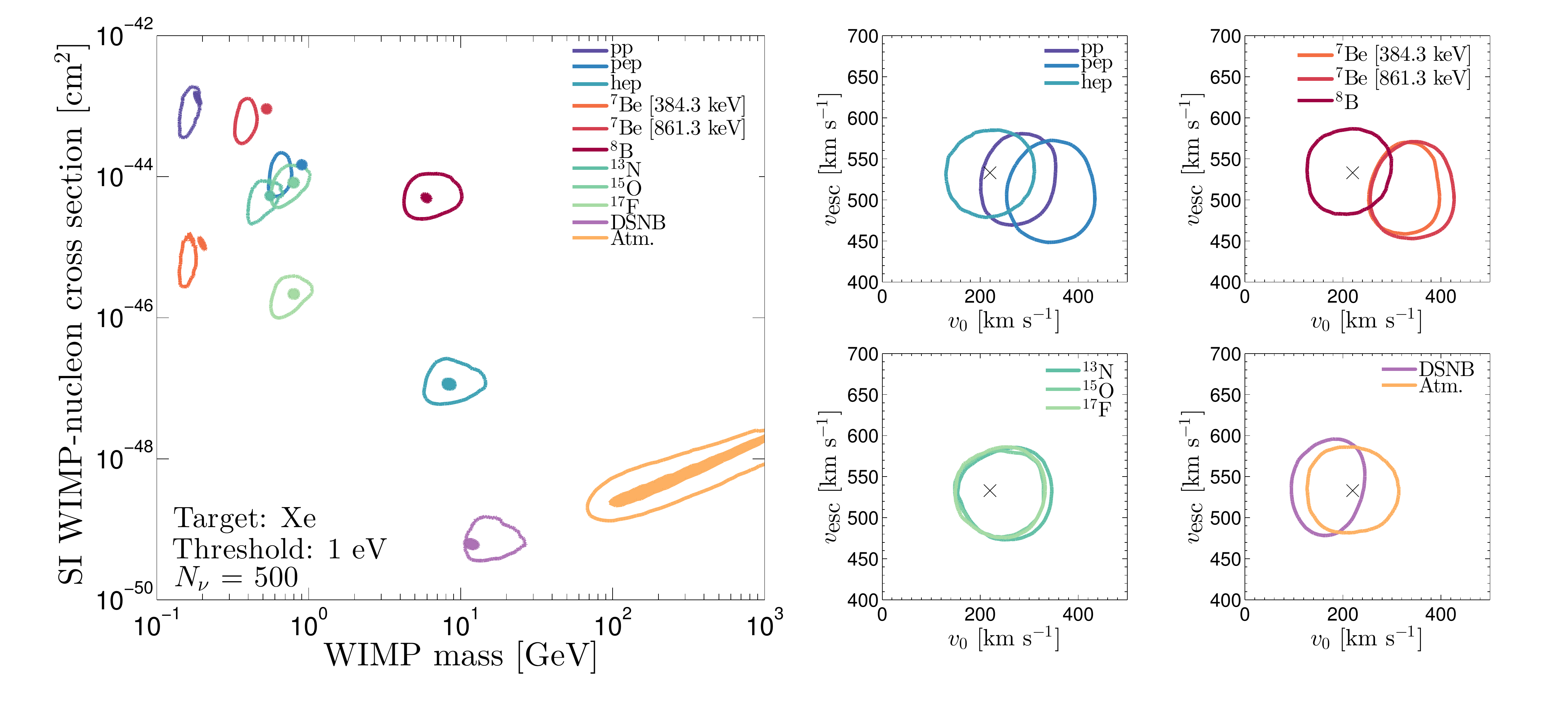}
\caption{{\bf Left:} Contours for the 95\% confidence region of the 2D marginalised posterior distribution in the WIMP mass-cross section plane under a WIMP-only hypothesis estimated for datasets made of each neutrino component individually. For each neutrino type the detector exposure was modified to give 500 expected events above a threshold of 1 eV. The filled contours correspond to the constraints with astrophysical parameters fixed and the unfilled contours are the same regions calculated when the astrophysical parameters are free and marginalised over. {\bf Right:} 95\% confidence contours in the $v_0-\vesc$ plane with WIMP mass and cross section marginalised over. The black cross shows the position of the fiducial SHM values $\vesc = 533$~km~s$^{-1}$ and $v_0 = 220$~km~s$^{-1}$.}
\label{fig:WIMPonly}
\end{center}
\end{figure*} 

The goal of this section is to demonstrate the effect neutrino backgrounds have on the measurement of WIMP parameters, both those of a particle physics and astrophysics origin. The discovery limits as derived in the previous section are a convenient way of showing how much of the WIMP mass-cross section parameter space is accessible to a given experiment. However they give us no infomation with regards how the other ingredient parameters of the WIMP signal may be constrained, which is undoubtedly a goal of direct detection experiments.

To reconstruct parameters using WIMP+neutrino data we will adopt a Bayesian approach. Following Bayes' theorem the posterior distribution which gives the probability distribution for parameters $\boldsymbol{\varphi}$ given a dataset $\mathcal{D}$, is
\begin{equation}
 \mathcal{P}(\boldsymbol{\varphi}|\mathcal{D}) = \frac{\mathscr{L}(\mathcal{D}|\boldsymbol{\varphi}) \pi(\boldsymbol{\varphi})}{\mathcal{Z}(\mathcal{D})} \, ,
\end{equation}
where $\mathcal{Z}$ is the Bayesian evidence, effectively a normalisation constant for our purposes. The probability distribution $\pi(\boldsymbol{\varphi})$ are the priors on each parameter, $\boldsymbol{\varphi} = \{m_\chi,\sigma_{\chi-n},\boldsymbol{\Phi},\boldsymbol{\Theta}\}$, and reflect our {\it a priori} knowledge of their true values. This distribution can be explored using nested sampling algorithms provided by the {\sc MultiNest} package~\cite{Feroz:2007kg,Feroz:2008xx,Feroz:2013hea}. A summary of the {\sc MultiNest} input specification used for parameter estimation is given in Table~\ref{tab:priors}. A disadvantage of the Bayesian approach in this context is that by only using a single dataset the stochastic fluctuations in any given one will influence the limits or constraints made using the posterior distribution. We can remove any potential bias due to statistical fluctuations in a single Monte-Carlo generated dataset by instead using an Asimov dataset where the observed data matches the theoretical prediction~\cite{Peter:2013aha}.
\begin{table}[h]
\ra{1.3}
\begin{tabular}{cccc}
\hline
\hline
\multirow{3}{*}{{\sc MultiNest}}& $N_\textrm{live}$& 2000& \\
				& tol & 0.001&\\
				& eff & 0.3&\\
\hline
\multirow{6}{*}{Priors} & $m_\chi$& log-flat & [0.1,1000] GeV \\
			& $\sigma_{\chi-n}$& log-flat & [10$^{-50}$,10$^{-30}$] cm$^2$ \\
			& $\rho_0$& Gaussian & $0.3 \pm 0.15$ GeV cm$^{-3}$ \\
			& $v_0$& Gaussian & $220 \pm 50$ km s$^{-1}$ \\
			& $\vesc$& Gaussian & $533 \pm 75$ km s$^{-1}$ \\
			& $\phi^j_\nu$ & Gaussian & [See Table~\ref{tab:neutrino} ] \\
\hline
\hline
\end{tabular}
\caption{Input specification and priors used for Bayesian parameter estimation using {\sc MultiNest}. }\label{tab:priors}
\end{table}

\subsection{WIMP only analysis}
\begin{figure*}[t]
\begin{center}
\includegraphics[trim = 0mm 0mm 0mm 0mm, clip,width=0.46 \textwidth,angle=0]{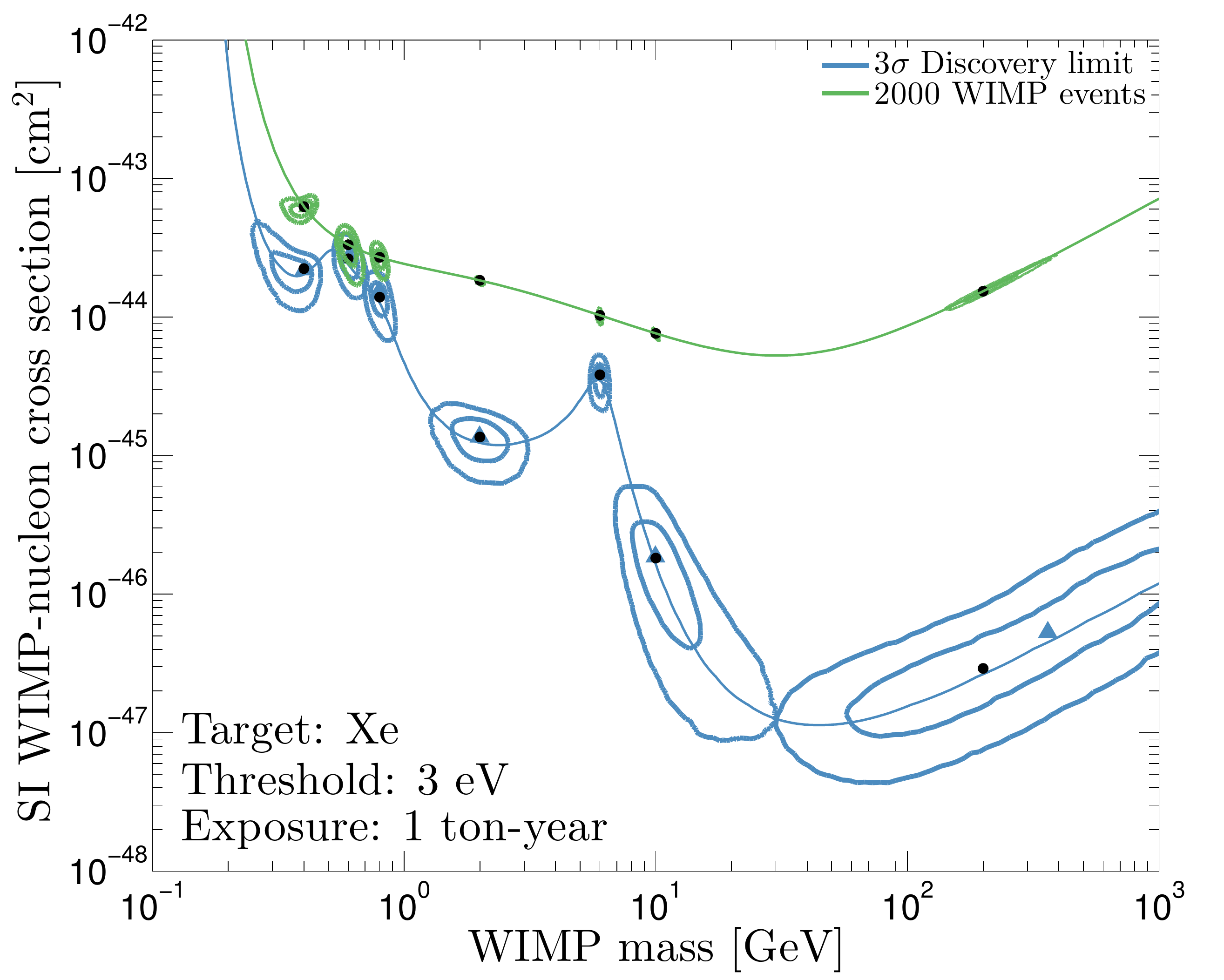}
\includegraphics[trim = 0mm 0mm 0mm 0mm, clip,width=0.46 \textwidth,angle=0]{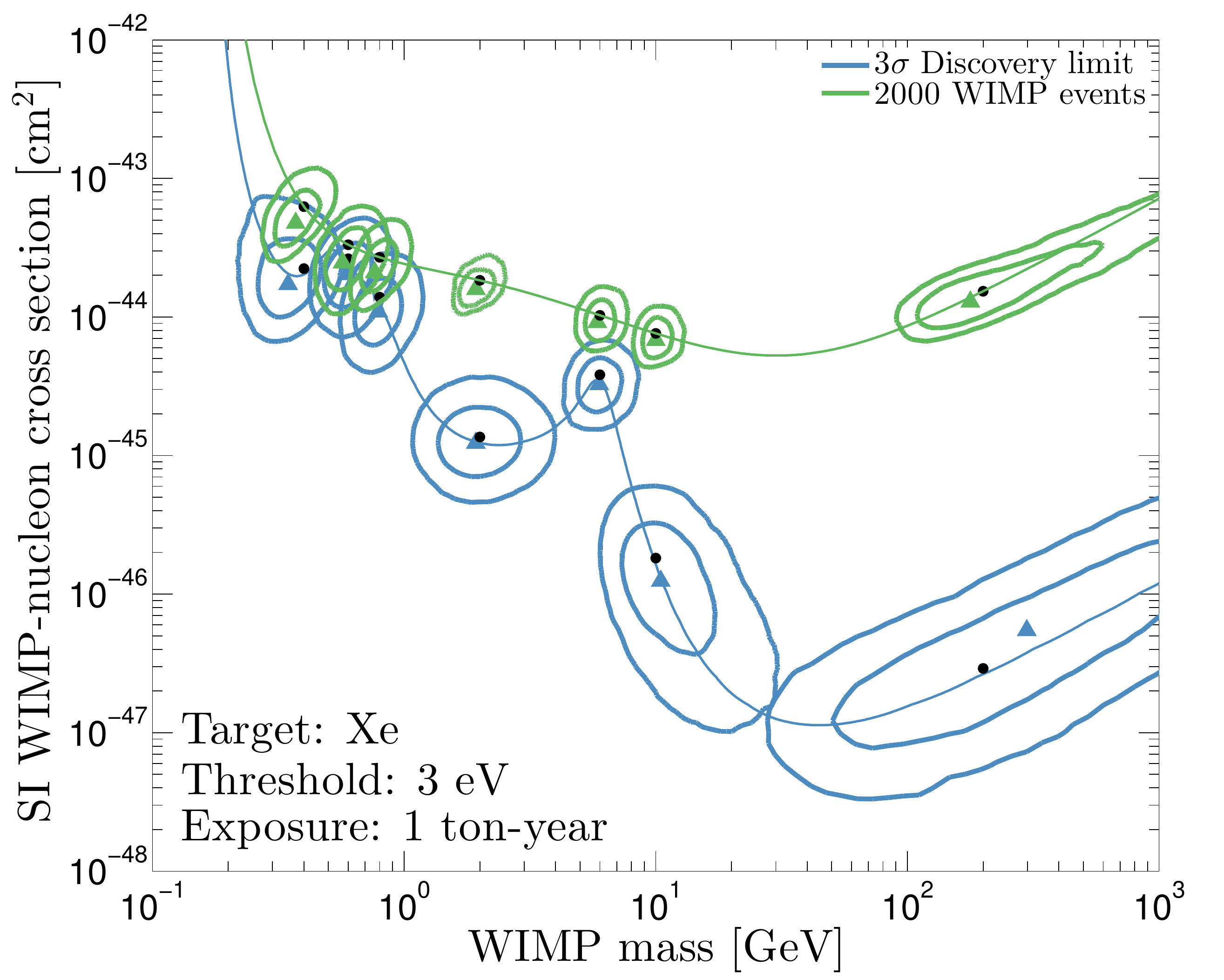}
\caption{{\bf Left:} Marginalised posterior distributions (68\% and 95\% CL) in the WIMP mass-cross section plane for a range of different input WIMP values (shown as black points). In each case the input value of cross section is chosen so that it either gives 2000 WIMP events (points along the green line) or lies just above the discovery limit for the experiment (blue line) as derived in Sec.~\ref{sec:nufloor}. The triangles show the location of maximum likelihood values. {\bf Right:} The same as the left hand plot, but with $v_0$, $\vesc$ and $\rho_0$ allowed to vary.} 
\label{fig:Parameter_reconstruction}
\end{center}
\end{figure*}

\begin{figure*}[t]
\begin{center}
\includegraphics[trim = 5mm 0 0mm 0mm, clip, width=0.49\textwidth,angle=0]{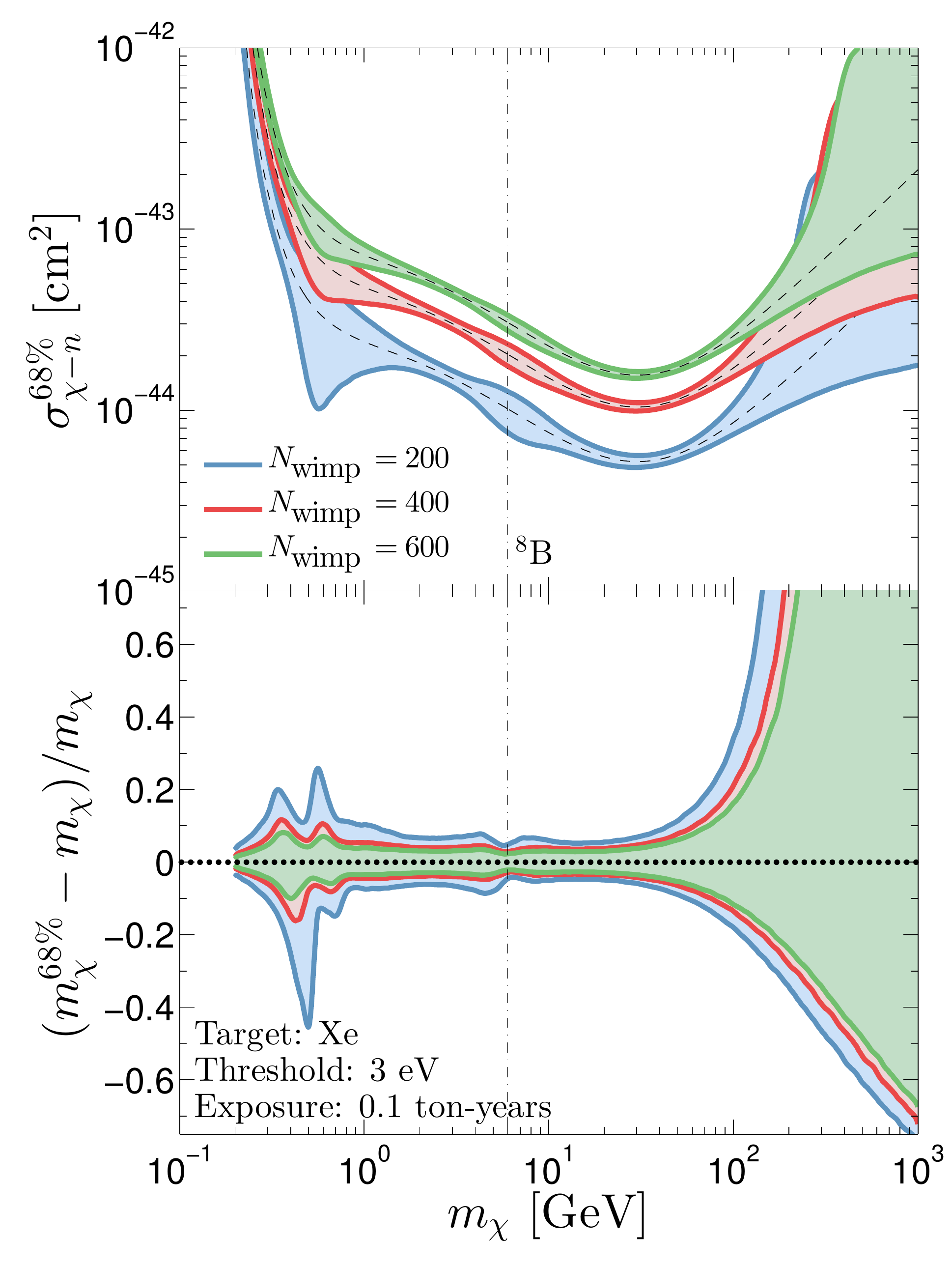}
\includegraphics[trim = 5mm 0 0mm 0mm, clip, width=0.49\textwidth,angle=0]{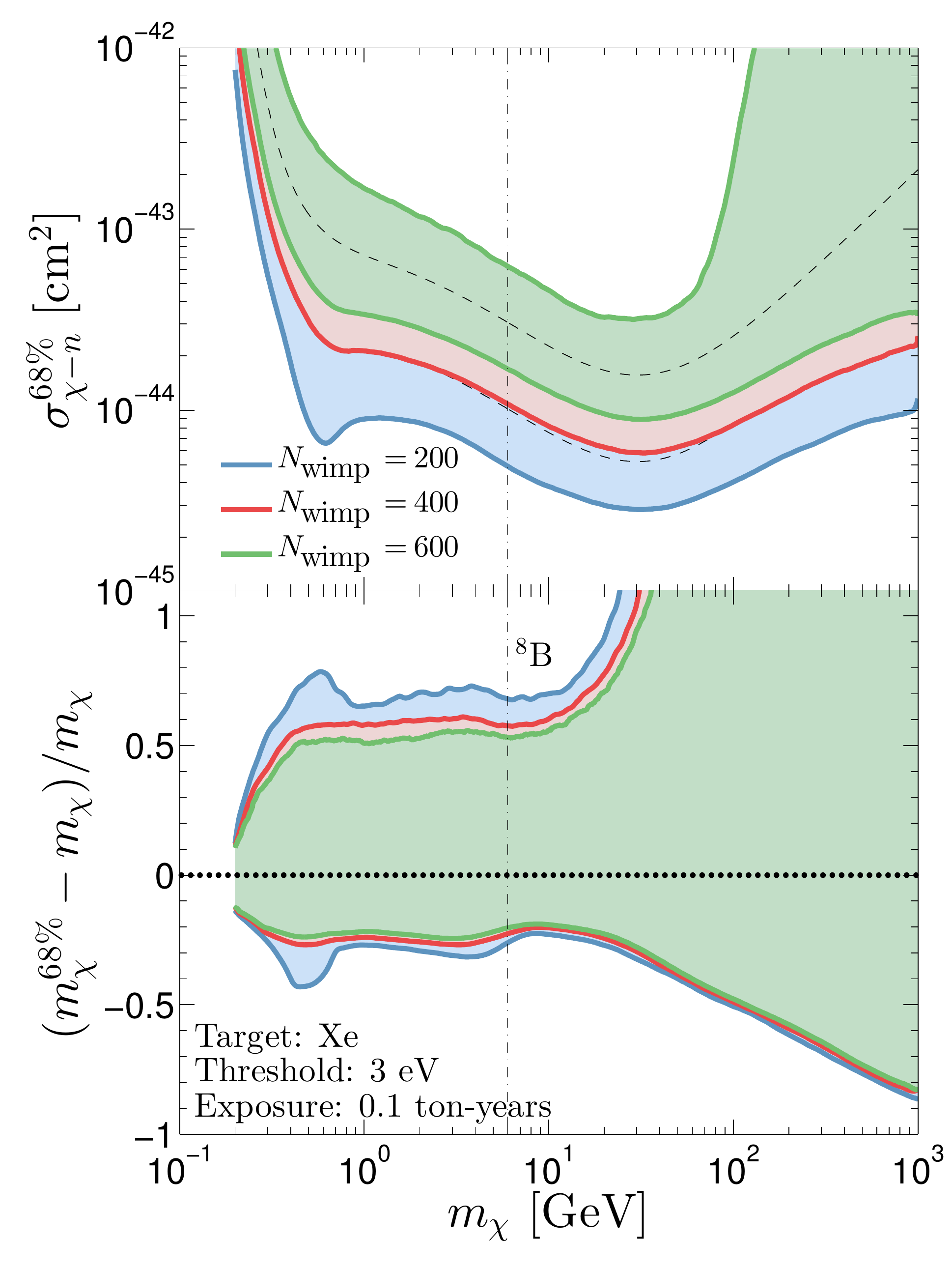}
\caption{{\bf Left:} Errors on the reconstructed values of $\sigma_{\chi-n}$ and $m_\chi$ as a function of input WIMP mass in the presence of the neutrino background. The coloured shaded regions enclose the 68\% profile likelihood errors on cross section $\sigma^{68/\%}_{\chi-n}$ (top panel) and WIMP mass $m^{68\%}_\chi$ (bottom panel, scaled by the input mass $m_\chi$). The input cross section for each WIMP mass is chosen so that the experiment observes 100 (blue), 300 (red) or 500 (green) WIMP events. The dashed lines in each region indicate those input values. The vertical dot-dashed lines indicate the mass for which WIMP and $^8$B neutrino recoil energies overlap (6 GeV). {\bf Right:} As in the left panel but with $v_0$, $\vesc$ and $\rho_0$ allowed to vary.}
\label{fig:error_recon}
\end{center}
\end{figure*} 

\begin{figure}[t]
\begin{center}
\includegraphics[trim = 0mm 0 0mm 0mm, clip, width=0.48\textwidth,angle=0]{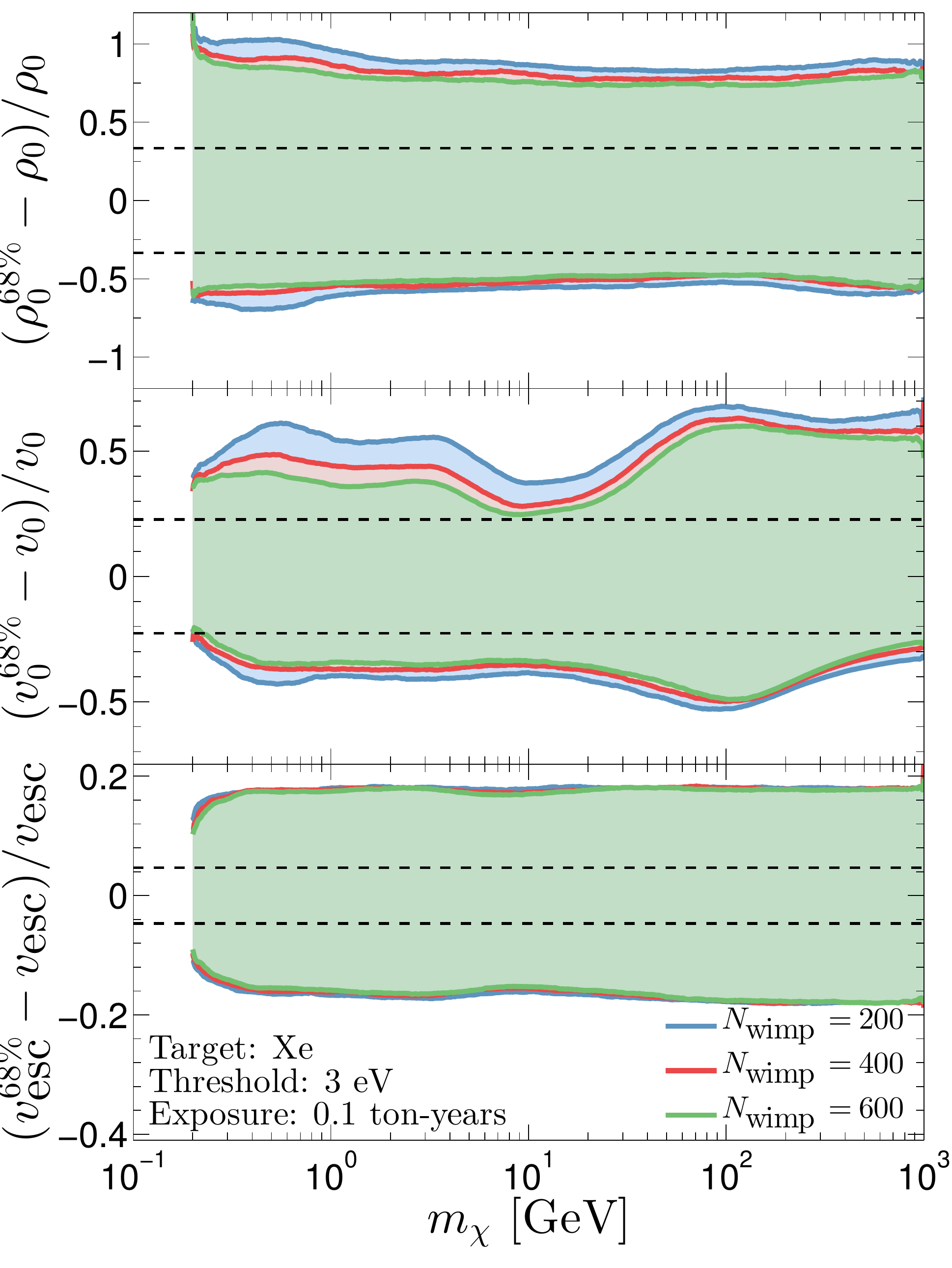}
\caption{Uncertainty in the reconstructed value of the three SHM parameters, from top to bottom, local density, Solar velocity and escape velocity. The shaded regions indicate the 68\% error bands calculated from the profile likelihood and are presented as a function of the input WIMP mass. For each value of WIMP mass the cross section is chosen to give a fixed value of WIMP events, either 200 (blue), 400 (red) or 600 (green). Also shown as dashed lines are the widths of the Gaussian prior on each parameter. The exposure chosen in this case, as in Fig.~\ref{fig:error_recon}, is 0.1 ton-years.}
\label{fig:error_recon_Maxwell_errs}
\end{center}
\end{figure} 
Before continuing with the complete WIMP+neutrino analysis we can gain insight into the influence of each source of neutrino on the experiment's overall sensitivity to the astrophysical parameters by performing a WIMP-only analysis on datasets comprised of a single neutrino source. The WIMP-only hypothesis consists of a likelihood of the form,
\begin{equation}
 \mathscr{L}(m_\chi,\sigma_{\chi-n},\boldsymbol{\Theta}) = \prod_{i=1}^{N_\textrm{obs}} \mathscr{P}(N^i_\textrm{bins} | N^i_\chi) \, .
\end{equation}
Where $N^i_\chi$ is the expected number of WIMP events, defined in Eq.~(\ref{eq:nchi}). 

In Fig.~\ref{fig:WIMPonly} we show the results of a WIMP-only analysis on each neutrino component individually. The exposure in each case was chosen so that the experiment observes 500 neutrino events above a threshold of 1 eV. In the left hand panel of Fig.~\ref{fig:WIMPonly} one can observe the WIMP masses that are fitted in the astrophysics fixed case (filled contours) match the positions of each peak in the neutrino floor seen in Fig.~\ref{fig:EL_vs_mass_allNeutrinos}. With astrophysics parameters, $\rho_0$, $\vesc$, $v_0$ unfixed, the interpreted WIMP parameters are quite different, as seen in the unfilled contours. The neutrino contributions when fitted under the this model that are least sensitive to changes in the astrophysical parameters appear to be $hep$, $^8$B, $^{13}$N, $^{15}$O, $^{17}$F and atmospheric neutrinos because the contours in the $m_\chi - \sigma_{\chi-n}$ plane do not change location when the astrophysical parameters are unfixed, and the contours in the $v_0-\vesc$ plane are centered on the fiducial values. The $^7$Be, $pp$ and $pep$ neutrinos seem to prefer slightly larger values of $v_0$ and the DSNB seems to fit to a slightly smaller value of $v_0$. These contours tell us how well each neutrino component mimics a WIMP signal. We can also observe that once astrophysical uncertainties are accounted for, the allowed range of cross section and WIMP mass becomes much larger, which is consistent with the result of Fig.~\ref{fig:DL_uncertainties}. 

One feature that we will note now as it will become important in subsequent results is the correlation between reconstructed WIMP mass and cross section for large values of $m_\chi$ ($>$100 GeV). This can be seen in the contours with atmospheric neutrino data. In this regime the energy dependence of WIMP event rate has a much weaker sensitivity to changes in the WIMP mass. At large WIMP mass the event rate is both proportional to cross section and inversely proportional to WIMP mass (due to the constant $\rho_0$, larger $m_\chi$ implies lower WIMP number density). These dependencies give rise to the positive correlation between WIMP mass and cross section in the contours at 100 GeV. This phenomenon has been explored in more detail in previous studies on the reconstruction of WIMP properties e.g., Refs.~\cite{Strigari:2009zb,Peter:2011eu,Peter:2013aha,Strege:2012kv}.

\subsection{Full analysis}
Following the WIMP only analysis we now return to the full WIMP+neutrino likelihood and discuss how the parameters of this model can be reconstructed using a Bayesian approach. The 2D marginalised posterior distributions in the WIMP mass-cross section plane for a range of input values are shown in Fig.~\ref{fig:Parameter_reconstruction} (a full set of 2D marginalised distributions for all parameters are given in the appendix~\ref{appendix1}). Here we choose a range of input WIMP masses and show the constraints recovered when the input cross section is large enough to produce 2000 WIMP events (green) or lies just above the discovery limit derived in Sec.~\ref{sec:nufloor} (blue). In both cases we use an exposure of 1 ton-year. When the WIMP mass lies just above the discovery limit the recovered constraints are very wide. Particularly for larger WIMP masses (above 200 GeV) when only 3-4 WIMP events are observed the constraints span over an order of magnitude in $m_\chi$ and $\sigma_{\chi-n}$. A feature that can be seen particularly in the constraints at 0.4, 10 and 200 GeV, is the 68\% and 95\% regions follow the shape of the discovery limit. This is because the discovery limit defines the region of the parameter space above which the WIMP events cannot be interpreted as fluctuations around the systematic uncertainty in the various neutrino fluxes. Hence when fitting the data to WIMP parameters the allowed regions should roughly follow the values permitted by the discovery limit. This effect can be seen in a slightly different way in the green contours which correspond to WIMP masses and cross sections that give 2000 WIMP events. Here the contours get noticably wider at low WIMP masses when the neutrino events mimic the WIMP signal.

Comparing the left and right hand panels of Fig.~\ref{fig:Parameter_reconstruction} we observe the effect of unfixing the values of the astrophysical parameters. We see a large increase in the size of the contours in both WIMP mass and cross section, particularly for light WIMPs ($<$ 10 GeV). As in the previous section we interpret this as being a result of the fact that when the value of $v_0$ is allowed to vary it allows a greater range of WIMP masses to explain recoil energies over a given energy range as well as a greater range of cross sections to explain the size of the event rate. The contours are also rounder and do not follow the shape of the discovery limit as closely, this is thanks to the freedom offered by a wider range of allowed parameter values to explain the data.

In Fig.~\ref{fig:error_recon} we show the 68\% error in the reconstructed values of WIMP mass and cross section calculated using the profile likelihood. In this case we show the dependence of this error over a range of WIMP masses from 0.1 to 1000 GeV where in each case the input cross section is chosen to produce a fixed number of WIMP events (100, 300 or 500). Here we see a similar effect to the upper contours in Fig.~\ref{fig:Parameter_reconstruction} but in greater detail. We can see now that at WIMP masses which match the recoil energy range of the underlying background, for instance 6 GeV and $^8$B neutrinos, the increase in uncertainty in cross section is coincident with a decrease in the uncertainty in WIMP mass. When $v_0$, $\rho_0$ and $\vesc$ are unfixed however both WIMP mass and cross section suffer an increase in error of up to a factor of 10 over the full range of input values. When these parameters are allowed to vary, less of the parameter space is mimicked by the background and hence there is more freedom in the allowed values, and correspondingly a larger uncertainty in the reconstructed $m_\chi$ and $\sigma_{\chi-n}$.

Finally in Fig.~\ref{fig:error_recon_Maxwell_errs} we show the same 68\% profile likelihood errors in the remaining WIMP parameters: $\rho_0$, $v_0$ and $\vesc$. For $\rho_0$ and $\vesc$ the error in their reconstructed values seems to only very marginally improve with increasing WIMP event number, $N_\textrm{wimp}$, as many more events than 600 are needed to significantly improve constraints on these parameters. Although around 0.2 GeV there is a rise in the error on $\rho_0$ as well as a slight decrease in the error on $\vesc$. The former can be understood to be a result of the sharp increase in the input cross section at very low WIMP masses, and the latter effect being a result of the recoil energies detectable at these masses being much more sensitive to the escape velocity. For the error on the reconstructed value of $v_0$ however we can observe an interesting dependency on the input value of $m_\chi$: rising around 100 GeV and reaching a minimum around 6 GeV.

In this section we have seen that the neutrino background has a detrimental effect on the potential WIMP parameter constraints that might be achievable with future detectors. At WIMP masses which are mimicked by individual neutrino backgrounds there is a large error in the reconstructed cross section induced by the systematic uncertainty in the neutrino flux. This problem remains when an estimation of the ingredient astrophysical parameters is attempted, and this is coupled with a large increase in the error in the reconstructed WIMP parameters even at masses which are not mimicked by neutrino backgrounds.

\section{Future detectors}\label{sec:future}
\begin{figure}[t]
\begin{center}
\includegraphics[trim = 0mm 0 0mm 0mm, clip, width=0.49\textwidth,angle=0]{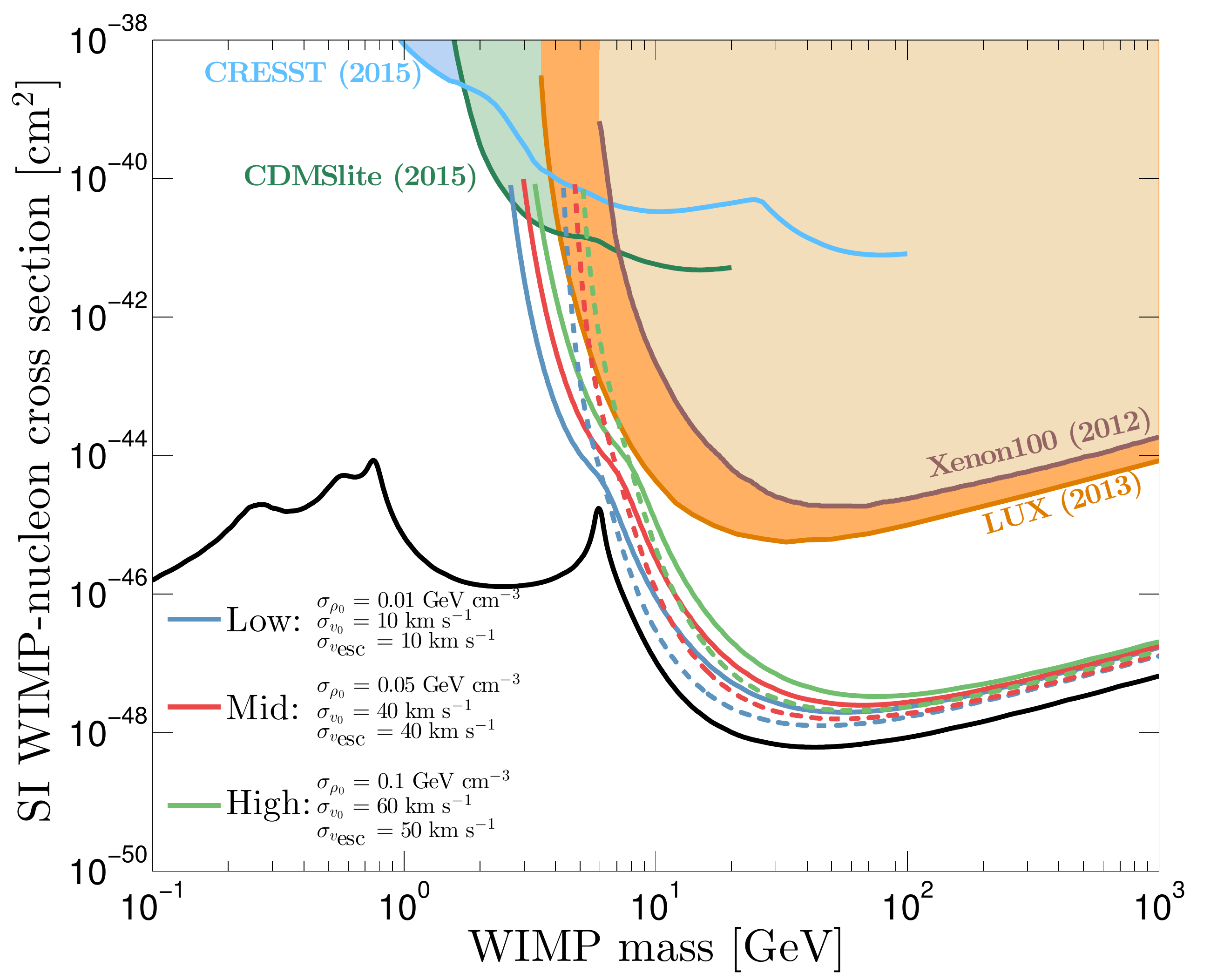}
\caption{Spin-independent WIMP nucleon cross section discovery limit as a function of WIMP mass for 3 sets of values of the uncertainty placed on the astrophysical parameters: local density, Solar velocity and escape velocity. The blue, red and green curves correspond to low, medium and high values for these uncertainties with 1$\sigma$ values shown. The solid lines are obtained when the recoil spectrum is convolved with a Gaussian energy resolution with $\sigma(E_r) = 0.8 E_r$ and the dashed lines when perfect energy resolution is considered. The filled regions are currently excluded by experiments, CRESST~\cite{Angloher:2015ewa}, CDMSlite~\cite{Agnese:2015nto}, Xenon100~\cite{Aprile:2012nq} and LUX~\cite{Akerib:2013tjd}. For comparison we show the full ($10^{-5}$ keV threshold) discovery limit indicated by the black line.}
\label{fig:DL_Resolution}
\end{center}
\end{figure}
Future direct detection experiments such as SuperCDMS~\cite{Agnese:2014aze}, Xenon1T~\cite{Aprile:2015uzo} and  LZ~\cite{Akerib:2015cja}, are poised to make the first detection of coherent neutrino-nucleus scattering. In this work so far we have used a very small threshold of 3 eV. This allowed us to map out the neutrino floor at low WIMP masses whilst keeping other experimental parameters such as the target nucleus and exposures consistent. One might argue however that this is an unrealistic expectation for future experiments, including those using a dual-phase liquid Xenon detector. The next generation of experiments are predicted to be sensitive to neutrino backgrounds with thresholds larger than those used in this work - once the energy resolution of the detector is taken into account. For example with a Xenon target, the maximum recoil energy from $^8$B neutrinos is 4.5 keV. However with an energy resolution of say 0.5 keV it is expected that some $^8$B Xenon recoils will leak into the energy sensitive window even with $E_th>4.5$~keV, provided $E_\textrm{th}$ isn't very much larger than 4.5 keV. 

The energy resolution is taken into account by convolving the event rate with a Gaussian resolution function defined by an energy dependent resolution $\sigma(E_r)$,
\begin{equation}
\frac{\textrm{d}R}{\textrm{d}E_r}(E_r) = \int_0^\infty \frac{1}{\sqrt{2\pi}\sigma(E_r)}e^{-\frac{(E_r-E_r')^2}{2\sigma^2(E_r)}}\frac{\textrm{d}R}{\textrm{d}E_r}(E_r')\textrm{d}E_r' \,.
\end{equation}

The results we will show in this section are for a 2 keV threshold Xenon detector with a 10 ton target mass over 1000 days running time, which is a reasonable approximation to what is expected in the near future beyond experiments such as LZ and Xenon1T. The energy resolution we take to be constant 80\% at 1$\sigma$ over the full energy range, i.e. $\sigma(E_r) = 0.8 E_r$. Figure~\ref{fig:DL_Resolution} shows the discovery limit for this detector when neutrinos and astrophysical uncertainties are included. We show results for three sets of values of the uncertainties and the limits when energy resolution is both included and ignored.

As previously mentioned, the finite energy resolution at low WIMP masses causes some recoils to leak into the energy sensitive window and hence the discovery limits below 10 GeV are lower than in the case of perfect resolution. This is interesting when neutrinos are included as it means that more $^8$B events are observed and the neutrino floor emerges around 6 GeV. As found in the results of Fig.~\ref{fig:DL_uncertainties} when the uncertainty on $v_0$ is larger, the floor appears at larger WIMP masses for the same reasons as discussed in Sec.~\ref{sec:astro}. In this case we see that for the largest uncertainties the discovery limits lie extremely close to the currently excluded region by LUX. The opposing effect of energy resolution however is that for larger WIMP masses some events are pushed {\it below} the energy threshold meaning that the discovery limit is raised to larger cross sections in this range, this can be seen for WIMP masses greater than 10 GeV. To reiterate the conclusion of Sec.~\ref{sec:astrounc} we state that the neutrino floor may be encountered by future direct detection experiments much sooner than previously thought. Even without the ultra-low thresholds used in previous sections, the neutrino floor still limits WIMP discovery when a finite energy resolution and modest threshold is considered. 

\begin{figure*}[t]
\begin{center}
\includegraphics[trim = 0mm 0mm 0mm 0mm, clip,width=\textwidth,angle=0]{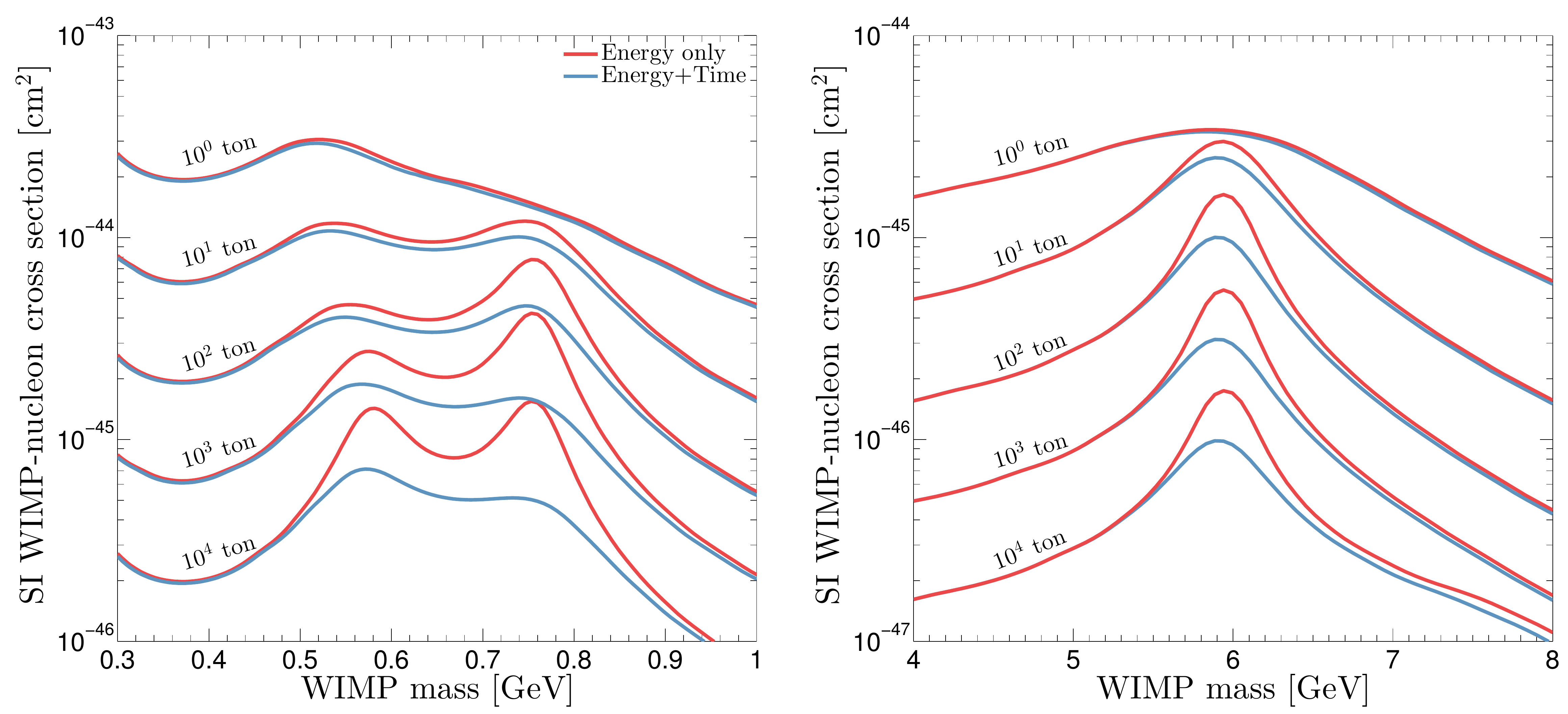}
\caption{Spin-independent WIMP nucleon discovery limits in the $0.3 - 1$~GeV (left) and $4 - 5$~GeV mass range (right). The red curves show the discovery limits obtained when only energy information is considered and the blue shows the improvement when time information is added. There are four sets of curves shown for 4 detector masses from 1 ton to 10$^4$ tons (top to bottom). In each case the exposure time was kept at a constant 1 year from the 1st of January 2016.} 
\label{fig:DL_Time}
\end{center}
\end{figure*}
\section{Including time information}\label{sec:time}
\begin{figure}[t]
\begin{center}
\includegraphics[trim = 0mm 0mm 0mm 0mm, clip,width=0.49\textwidth,angle=0]{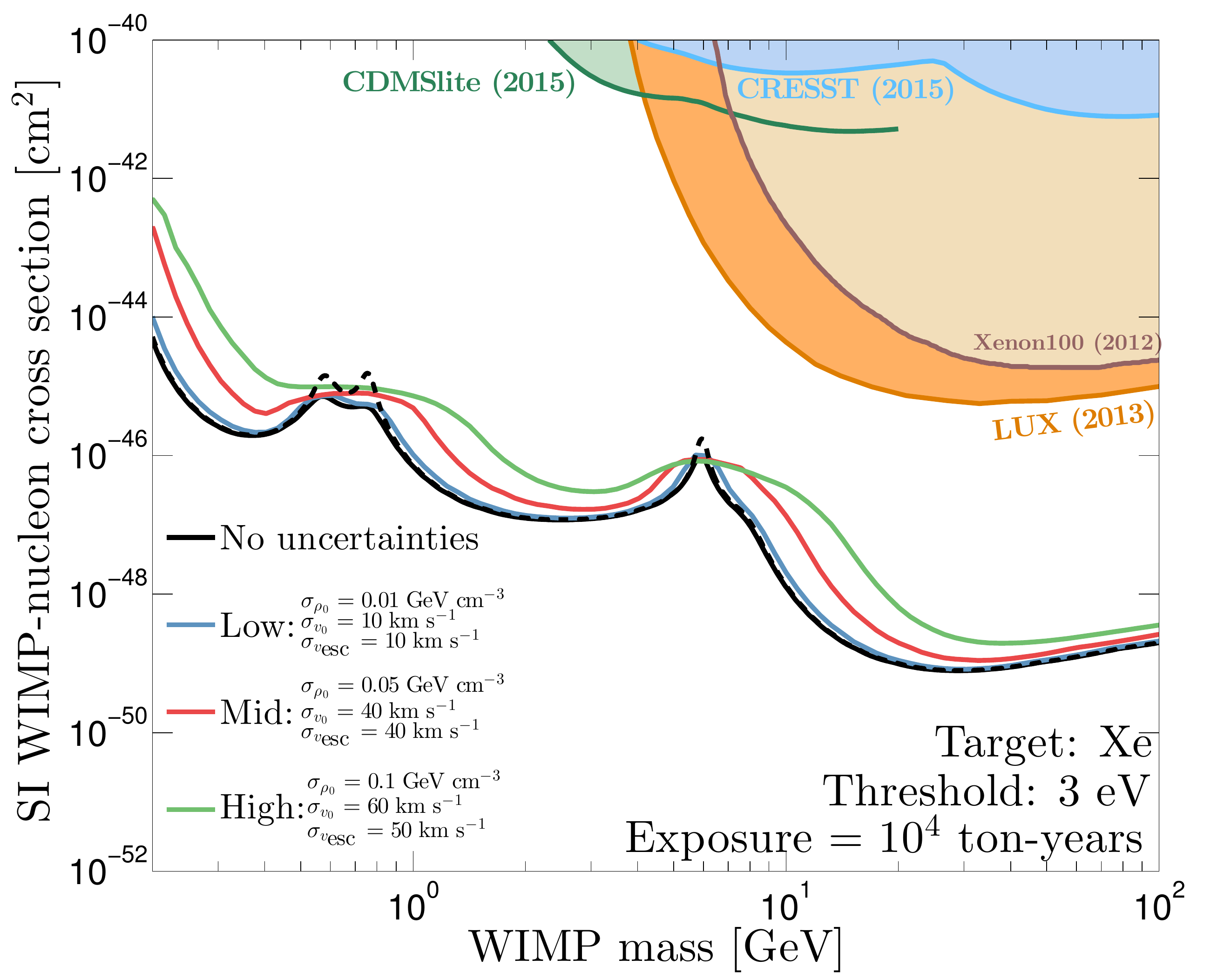}
\caption{Spin-independent discovery limit as a function of WIMP mass calculated with the inclusion of astrophysical uncertainties and time dependence in the profile likelihood analysis. The blue, red and green curves correspond to 3 sets of 1$\sigma$ uncertainties on the parameters $\rho_0$, $v_0$ and $\vesc$. The size of the uncertainties are labelled from low to high with values indicated. The black lines are the discovery limits without the inclusion of astrophysical uncertainties: with and without the inclusion of time dependence (solid and black lines respectively). The filled regions are currently excluded by experiments, CRESST~\cite{Angloher:2015ewa}, CDMSlite~\cite{Agnese:2015nto}, Xenon100~\cite{Aprile:2012nq} and LUX~\cite{Akerib:2013tjd}.} 
\label{fig:DL_Time_astro}
\end{center}
\end{figure}
It was shown in Ref.~\cite{Davis:2014ama} that the time dependence of the WIMP and Solar neutrino event rates provides a distinguishing feature between the two signals which can help circumvent the neutrino floor. The WIMP signal is time dependent because of a well known annual modulation effect due to the motion of the Earth with respect to the Sun~\cite{annual}. We can insert this time dependence into the WIMP calculation by simply including the additonal velocity $\textbf{v}_\textrm{EarthRev}(t)$ in the lab velocity $\textbf{v}_\textrm{lab}$. Details on the time dependence of this velocity component can be found in Ref.~\cite{daily}. 

The Solar neutrino flux also exhibits an annual modulation in the event rate due to the eccentricity of the Earth's orbit which causes a change in the Earth-Sun distance over the course of a year. The time dependence can be written,
\begin{equation}
  \frac{\textrm{d} \Phi (t) }{\textrm{d}E_\nu}  =  \frac{\textrm{d} \Phi}{\textrm{d} E_\nu}\left[ 1 + 2\epsilon\cos\left(\frac{2\pi(t- t_\nu)}{T_\nu}\right) \right] \,,
\label{eq:solarneutrinoflux}
\end{equation}
where $t$ is the time from January 1st, $\epsilon = 0.016722$ is the eccentricity of the Earth's orbit, $t_{\nu} = 3$ days is the time at which the Earth-Sun distance is shortest (and hence the Solar neutrino flux is largest) and $T_{\nu} = 1$ year. Both the Solar neutrino and WIMP event rates have a $\sim 5$\% annual modulation but they peak at times separated by about 5 months.

We now perform the energy+time analysis by extending the likelihood,
\begin{align}
 \mathscr{L}(m_\chi,\sigma_{\chi-n},\boldsymbol{\Phi},\boldsymbol{\Theta}) =& \prod_{i=1}^{N_{E_r}}\prod_{j=1}^{N_t} \mathscr{P} \left(N_\textrm{obs}^i \bigg| N^i_\chi + \sum_{j=1}^{n_\nu} N^{ij}_\nu(\phi^k)\right) \nonumber \\ 
& \times \prod_{k=1}^{n_\nu} \mathcal{L}(\phi^k) \nonumber \\
& \times \prod_{l=1}^{n_\theta} \mathcal{L}(\theta^l) \, .
\end{align}
Where now we bin in both energy and time with $N_{E_r}$ and $N_t$ bins respectively. The number of WIMP events in bin $(i,j)$ is,
\begin{equation}
N_\chi^{ij}(m_\chi,\sigma_{\chi-n},\boldsymbol{\Theta}) =\mathcal{M}\int_{E_{i}}^{E_{i+1}}\int_{t_j}^{t_{j+1}} \frac{\textrm{d}R_\chi(t)}{\textrm{d}E_r} \textrm{d}t\, \textrm{d}E_r \, ,
\end{equation}
and $N_\nu^{ij}(\phi^k)$ is the number of expected neutrino events from the $k$th neutrino species,
\begin{equation}
N^{ij}_\nu (\phi^k) = \mathcal{M}\int_{E_{i}}^{E_{i+1}}\int_{t_j}^{t_{j+1}} \frac{\textrm{d}R_\nu(t)}{\textrm{d}E_r}(\phi^k) \textrm{d}t\, \textrm{d}E_r \, ,
\end{equation}
where $\mathcal{M}$ is the mass of the detector.

In Fig.~\ref{fig:DL_Time} we show the the improvement on the discovery limits obtained for WIMP masses which are mimicked best by Solar neutrino backgrounds. The improvement is most noticeable between 0.4-1 GeV when the $^7$Be, $pep$, $^{13}$N, $^{15}$O and $^{17}$F neutrinos play the biggest role, as well as at 6 GeV when the floor is induced by $^8$B neutrinos. Outside of these specific mass ranges when the event rate energy dependencies do not overlap the improvement offered by time information is very small. Above 10 GeV for example (for exposures with a very small atmospheric and supernovae background rates) the discovery limit cross section is simply set by Poisson statistics at a value which produces a sufficient number of WIMP events to be significant over the systematic uncertainty on the total background flux. Moreover, because the annual modulation amplitudes are small, to observe the benefit of time information one needs to go to large exposures which see in excess of $\mathcal{O}(1000)$ neutrino events.

Incorporating uncertainties on the SHM parameters into the energy+time analysis we obtain limits shown in Fig.~\ref{fig:DL_Time_astro}. The results we obtain here are analogous to those of the Fig.~\ref{fig:DL_uncertainties} only now that we are working with a larger exposure which sees a number of events sufficient to discriminate neutrinos and WIMPs using their respective time dependencies. Again the discovery limit under the assumption of the largest values of uncertainty is up to an order of magnitude higher than the astrophysics fixed case. However it still remains below the energy only limit around the peaks due to the Solar neutrino contributions. As we have performed this analysis at a very large detector mass of $10^4$ tons, we do not see the same proximity to the LUX region as in Figs.~\ref{fig:DL_uncertainties} and~\ref{fig:DL_Time_astro}. For smaller exposures, closer to those used in previous results, we do not expect any extra discrimination power with the addition of time information.

\section{Summary}\label{sec:conc}
In this work we have demonstrated the impact of astrophysics uncertainties in the analysis of WIMP and neutrino data. We have used two different statistical approaches to both evaluate the discoverability of a WIMP signal by a given experiment as well as demonstrating the ability of an experiment to measure the underlying parameters of the WIMP and neutrino signals. Relaxing the assumption of perfectly known astrophysics parameters such as the Solar velocity, escape speed and local WIMP density results in a shift in the shape of the neutrino floor as a function of WIMP mass and cross section. We find that if there are reasonably large uncertainties in the various astrophysics parameters (close to those currently known) then neutrinos will become important to future direct detection experiments sooner than previously realised. 

When attempting to reconstruct the input WIMP and neutrino parameters we find that unfixing the astrophysics parameters induces a significant increase in the uncertainty of the reconstruction. When the astrophysical parameters are fixed and only WIMP mass and cross section are reconstructed we see that for input WIMP masses which suffer the most overlap in recoil energies with different neutrino contributions there is trade off between a small error in mass with a large error in reconstructed cross section. This error in cross section is almost entirely controlled by the systematic uncertainty in the relevant neutrino flux, whereas the error in WIMP mass is small because of the similarity between the recoil energy ranges of WIMP and neutrino induced recoils. For this reason then we see that including uncertainty in the value of the astrophysical parameters, which in turn induces an uncertainty in the allowed range of recoil energies, there is a large increase in the error of the reconstructed WIMP mass. For the remaining parameters we see that the error on the reconstructed values has little dependence on the input WIMP parameters, with only $v_0$ having a noticeable relationship by being the parameter that most affects the shape of the WIMP recoil spectrum.

The first detection of coherent neutrino-nucleus scattering is expected to be made with the forthcoming generation of ton-scale direct detection experiments~\cite{Akerib:2015cja}. When this occurs it will be crucial to begin to implement strategies for dealing with neutrino backgrounds. This can be achieved in a number ways. As can be seen in this work, as well as that of Ref.~\cite{Davis:2014ama} the number of events observed at these detector masses are not yet enough to utilise the time dependence of the WIMP and neutrino signals to discriminate the two. For spin-dependent interactions as well as non-relativistic effective field theory operators, complementarity between target materials will be a powerful and relatively easy method for discriminating neutrinos~\cite{neutrinoRuppin,Dent:2016iht}. Independent of the WIMP-nucleus interaction however, directional detection, if experimentally feasible, will prove the most powerful scheme for distinguishing WIMPs and neutrinos~\cite{Grothaus:2014hja, O'Hare:2015mda}. The angular signature of WIMP and neutrino recoils are entirely distinct and this is true for any relationship set of astrophysical inputs or WIMP-nucleus interactions. However for the upcoming generation of direct detection experiments which will lack sensitivity to either direction or time dependence, we have shown that a better understanding of the uncertainty in the astrophysical dependence of a predicted WIMP signal will be vital to understand in order to deal with the neutrino background.

\begin{acknowledgments}
The author thanks Anne M. Green for helpful discussion and comments. The author is supported by the Science \& Technology Facilities Council (STFC).
\end{acknowledgments}

\appendix
\section{Full constraints}\label{appendix1}
\begin{figure*}[t]
\begin{center}
\includegraphics[trim = 0mm 0 0mm 0mm, clip, width=\textwidth,angle=0]{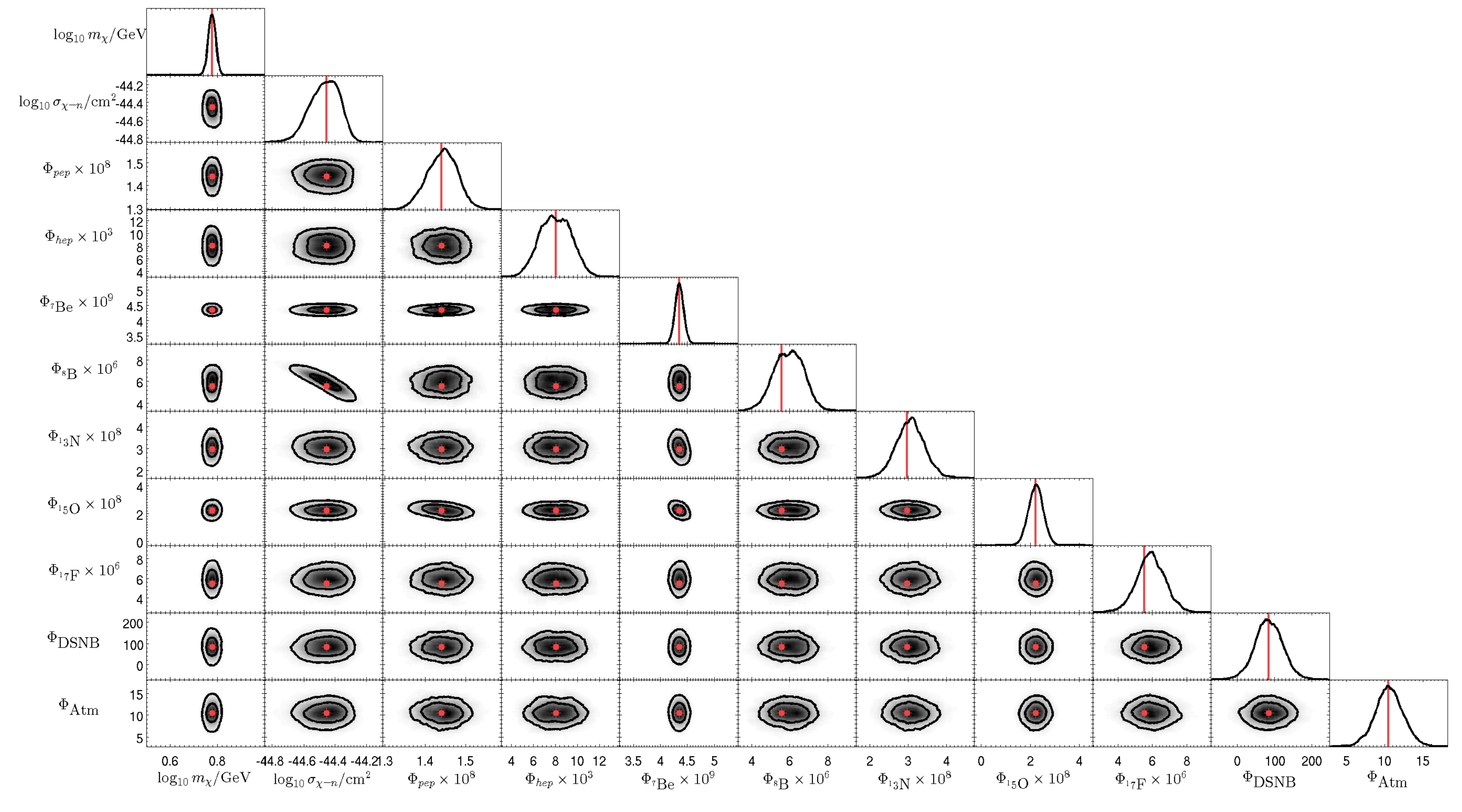}
\includegraphics[trim = 0mm 00mm 00mm 0mm, clip, width=\textwidth,angle=0]{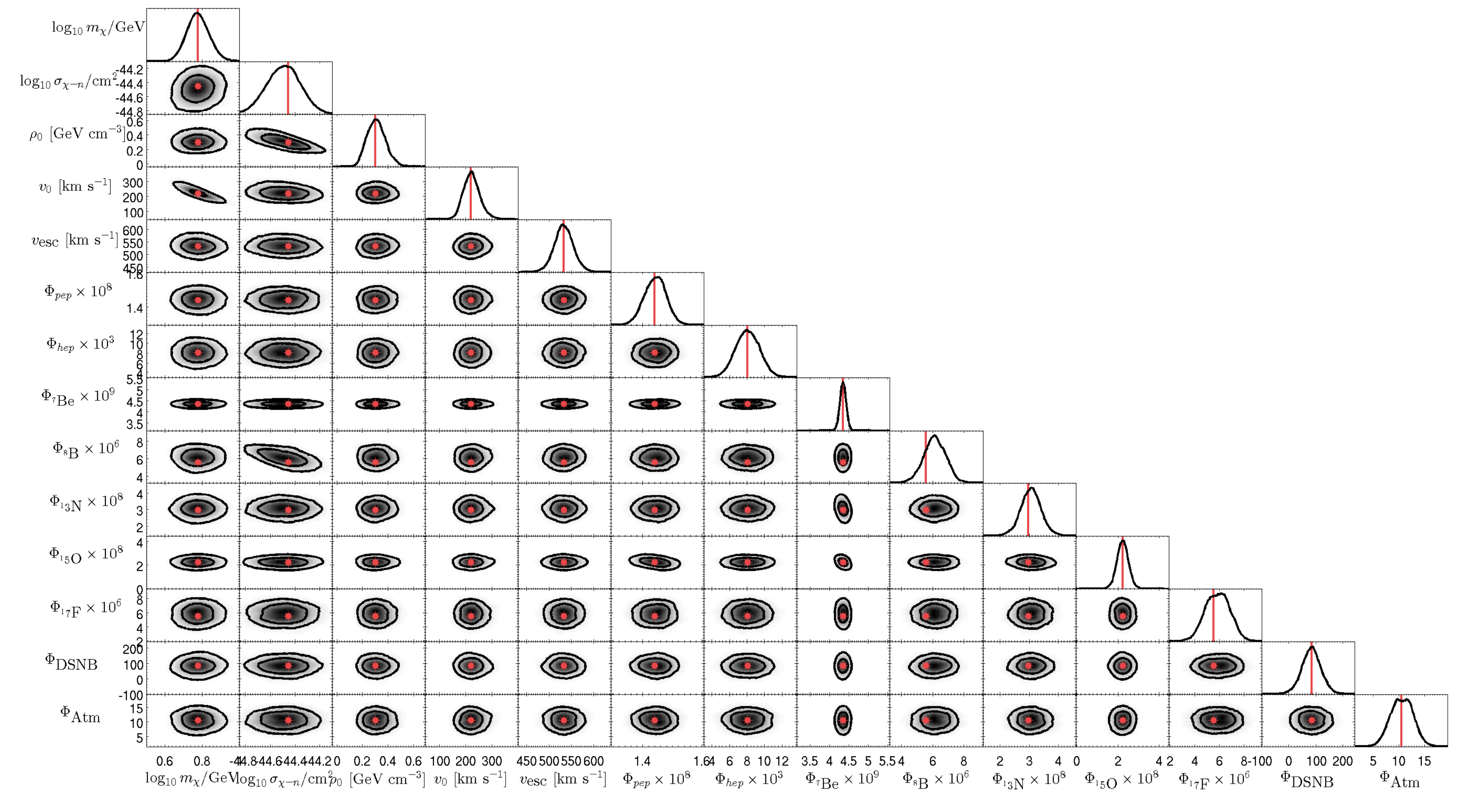}
\caption{{\bf Top:} Triangle plot of the posterior distribution of the WIMP+neutrino model. The off diagonal plots show the 2D marginalised posterior distribitions and the diagonal entries show the 1D distributions. The red lines/points indicate the input value(s) of each parameter. The posterior distribution was calculated using the {\sc MultiNest} nested sampling algorithm with input specification and prior ranges given in Table~\ref{tab:priors}. The neutrino fluxes $\Phi_{\nu_i}$ are in units of MeV cm$^{-2}$ s$^{-1}$. {\bf Bottom:} Same triangle plot of the 2D and 1D posterior distributions but including uncertainties on the parameters of the SHM: $v_0$, $\vesc$ and $\rho_0$.}
\label{fig:Parameter_reconstruction_triangle}
\end{center}
\end{figure*} 

In this appendix we show the full 2D marginalised constraints on each parameter in both the initial case with the astrophysical parameters fixed (top of Fig.~\ref{fig:Parameter_reconstruction_triangle}), and then in the case where we allow the parameters to vary under a Gaussian prior (bottom of Fig.~\ref{fig:Parameter_reconstruction_triangle}). The prior ranges are shown in Table~\ref{tab:priors}. In both cases we use Asimov data as a representative to remove the dependence on the stochastic nature of an individual dataset, however we do not expect there to much dependence on this choice due to the large number of events observed in this example. For the WIMP mass we pick 6 GeV with a cross section that lies just above the discovery limit so there is maximum overlap between the WIMP signal and the $^8$B signal. In both cases the majority of the parameters are recovered accurately and with little to no correlation between them. There is one notable exception in the astrophysics fixed case between cross section and the $^8$B flux, as is to be expected when the signal and background event rates are extremely similar. In the second case however we note two more additional correlations, a degeneracy between $\rho_0$ and $\sigma_{\chi-n}$ because they both appear as coefficients in the WIMP event rate and a negative correlation between $v_0$ and $m_\chi$ because WIMP event number is both proportional to $v_0$ and inversely proportional to $m_\chi$ at low WIMP masses (i.e., with recoil energies close to the threshold).

\end{document}